%% file: main_arxiv.tex
\documentclass[twocolumn,aps,pra,showpacs,10pt]{revtex4-2}
\usepackage{graphicx,amsmath,amssymb} 
\usepackage{mathtools}
\usepackage{enumerate}
\DeclarePairedDelimiter\bra{\langle}{\rvert}
\DeclarePairedDelimiter\ket{\lvert}{\rangle}
\DeclarePairedDelimiterX\braket[2]{\langle}{\rangle}{#1\,\delimsize\vert\,\mathopen{}#2}

\usepackage[colorlinks=true,	linkcolor=blue,urlcolor=blue,anchorcolor=blue,citecolor=blue,bookmarksnumbered]{hyperref}

\begin{document}

\input{main.tex}
\onecolumngrid
\appendix

\section*{Supplemental Material}
\setcounter{equation}{0}
\setcounter{figure}{0}
\setcounter{table}{0}

\renewcommand{\theequation}{S\arabic{equation}}
\renewcommand{\thefigure}{S\arabic{figure}}
\renewcommand{\thetable}{S\Roman{table}}

\makeatletter
\@removefromreset{equation}{section}
\@removefromreset{figure}{section}
\makeatother

\input{SM.tex}   
\twocolumngrid

\input{main_arxiv.bbl}
\end{document}

%% file: main.tex
%
%
%
\title{Reciprocal Floquet thermalization in a one-dimensional Rydberg atom array}
\author{Yunhui He$^{1,2}$}
\author{Yuechun Jiao$^{1,2}$}
\email{ycjiao@sxu.edu.cn}
\author{Jianming Zhao$^{1,2}$}
\author{Weibin Li$^{3}$}
\email{weibin.li@nottingham.ac.uk}
\affiliation{$^1$State Key Laboratory of Quantum Optics Technologies and Devices, Institute of Laser Spectroscopy, Shanxi University, Taiyuan 030006, China\\
$^2$Collaborative Innovation Center of Extreme Optics, Shanxi University, Taiyuan 030006, China\\
$^3$School of Physics and Astronomy, and Centre for the Mathematics and Theoretical Physics of Quantum Non-equilibrium Systems, University of Nottingham, Nottingham, NG7 2RD, United Kingdom}
\begin{abstract}
Periodically driven Floquet quantum systems hold great promise for engineering exotic quantum phases and matter, but are often limited by rapid thermalization. In this work, we propose and demonstrate a square-wave-modulated Floquet engineering protocol to steer and study the thermalization dynamics in one-dimensional Rydberg atom arrays. We identify a reciprocal Floquet thermalization mechanism, which is triggered when the combination of laser detuning and Rydberg atom interactions inversely matches the Floquet period. The level statistics show narrow peaks when the reciprocal condition is met, while thermalization is suppressed between two adjacent peaks. We extract signatures of thermalization and its suppression from the stroboscopic evolution of the atomic population. Critically, thermalization occurs in a disorder-free regime, with rapid equilibration achieved within the Rydberg lifetime and experimentally accessible initial states. Our study establishes a robust framework for exploring thermalization-to-localization transitions and designing effective Hamiltonians, and highlights the unique potential of the Rydberg atom array setting for quantum simulations.
\end{abstract}

\maketitle
\emph{Introduction---}
Floquet engineering~\cite{abaninExponentiallySlowHeating2015a,moriRigorousBoundEnergy2016,abaninRigorousTheoryManyBody2017,eckardtColloquiumAtomicQuantum2017,okaFloquetEngineeringQuantum2019,deckerFloquetEngineeringTopological2020,borishTransverseFieldIsingDynamics2020,geier2021floquet,weitenbergTailoringQuantumGases2021,sunEngineeringProbingNonAbelian2023,zhang2025suppressing} enables the realization of effective Hamiltonians and many-body phases that are often inaccessible in static Hamiltonians~\cite{nandkishore2015manybody,bukovUniversalHighfrequencyBehavior2015,kuwaharaFloquetMagnusTheory2016}. This gives an essential tool for creating quantum matter~\cite{struckEngineeringIsingXYSpinmodels2013,aidelsburgerRealizationHofstadterHamiltonian2013,goldman2014periodically,cooperTopologicalBandsUltracold2019,leonardRealizationFractionalQuantum2023} and exploring many-body dynamics~\cite{russomanno2012periodic,zhang2016floquet,jurcevic2017direct} and phases~\cite{khemani2016phase,else2016floquet,zhang2017observation,Kyprianidis2021Observation,moessnerEquilibrationOrderQuantum2017,yeFloquetPhasesMatter2021}. However, in isolated quantum systems, energy absorption from the Floquet driving  often leads to heating and rapid thermalization to an effective infinite temperature thermal state~\cite{dalessio2014longtimea, lazaridesEquilibriumStatesGeneric2014,weidingerFloquetPrethermalizationRegimes2017,kimTestingWhetherAll2014,haldar2018onset,fleckenstein2021thermalization,morningstarUniversalityClassesThermalization2023a,dagManybodyQuantumChaos2023,mukherjeeFloquetThermalizationInstantonsDynamical2024a}, which has been observed experimentally~\cite{choiProbingQuantumThermalization2019,pengFloquetPrethermalizationDipolar2021b,martinControllingLocalThermalization2023,lei2025quantum}. Thermalization washes out correlations and suppresses nontrivial dynamics, limiting the utility of Floquet engineering~\cite{jensenStatisticalBehaviorDeterministic1985,deutsch1991quantuma,srednicki1994chaos,rigolThermalizationItsMechanism2008b}. Controlling and preventing heating, through techniques such as many-body localization~\cite{ponte2015manybody,lazaridesFateManyBodyLocalization2015,khemani2016phase,zhang2016floquet,abaninTheoryManybodyLocalization2016,bordiaPeriodicallyDrivingManybody2017,deckerFloquetEngineeringTopological2020,morningstarAvalanchesManybodyResonances2022,yousefjaniFloquetinducedLocalizationLongrange2023a,sierantStabilityManybodyLocalization2023,mukherjeeArrestingQuantumChaos2025}, dynamical localization~\cite{jiSuppressionHeatingQuantum2018,heylQuantumLocalizationBounds2019,sieberer2019digital}, and high-frequency  driving~\cite{abaninExponentiallySlowHeating2015a,bukovUniversalHighfrequencyBehavior2015,kuwaharaFloquetMagnusTheory2016,abaninEffectiveHamiltoniansPrethermalization2017a}, is important for the study of Floquet-driven quantum many-body systems.

Remarkable developments in neutral atom tweezer arrays have enabled the manipulation of long-range Rydberg interactions via sinusoidal laser detuning~\cite{zhao2023floquettailored,basak2018periodically,mallavarapuPopulationTrappingPair2021}. This Floquet engineering not only facilitates the study of exotic thermalization within restricted Hilbert spaces~\cite{zhao2025observation} but also provides a framework that is important for quantum simulation~\cite{nguyenQuantumSimulationCircular2018,turner2018weak,bluvstein2021controlling,ebadiQuantumPhasesMatter2021,schollQuantumSimulation2D2021, geier2021floquet} and quantum computation~\cite{saffmanQuantumComputingAtomic2016,levineParallelImplementationHighFidelity2019,bluvsteinQuantumProcessorBased2022,grahamMultiqubitEntanglementAlgorithms2022,shaoRydbergSuperatomsArtificial2024a,bluvsteinLogicalQuantumProcessor2024, wuQuantumComputationFloquet2025a} with the Rydberg atom array setting~\cite{browaeys2020manybody}. Despite growing interest and potentials~\cite{mukherjee2020dynamics,sugiuraManybodyScarState2021,mukherjeePeriodicallyDrivenRydberg2022,ghosh2023detecting,nishadQuantumSimulationGeneric2023,martinsQuasiperiodicFloquetGibbsStates2025,tianEngineeringFrustratedRydberg2025,koyluoglu2025floquet}, protocols to directly manipulate and probe Floquet thermalization in the Rydberg atom arrays remain largely unexplored.

\begin{figure}[htp!]
	\centering  
	\includegraphics[width=0.95\linewidth]{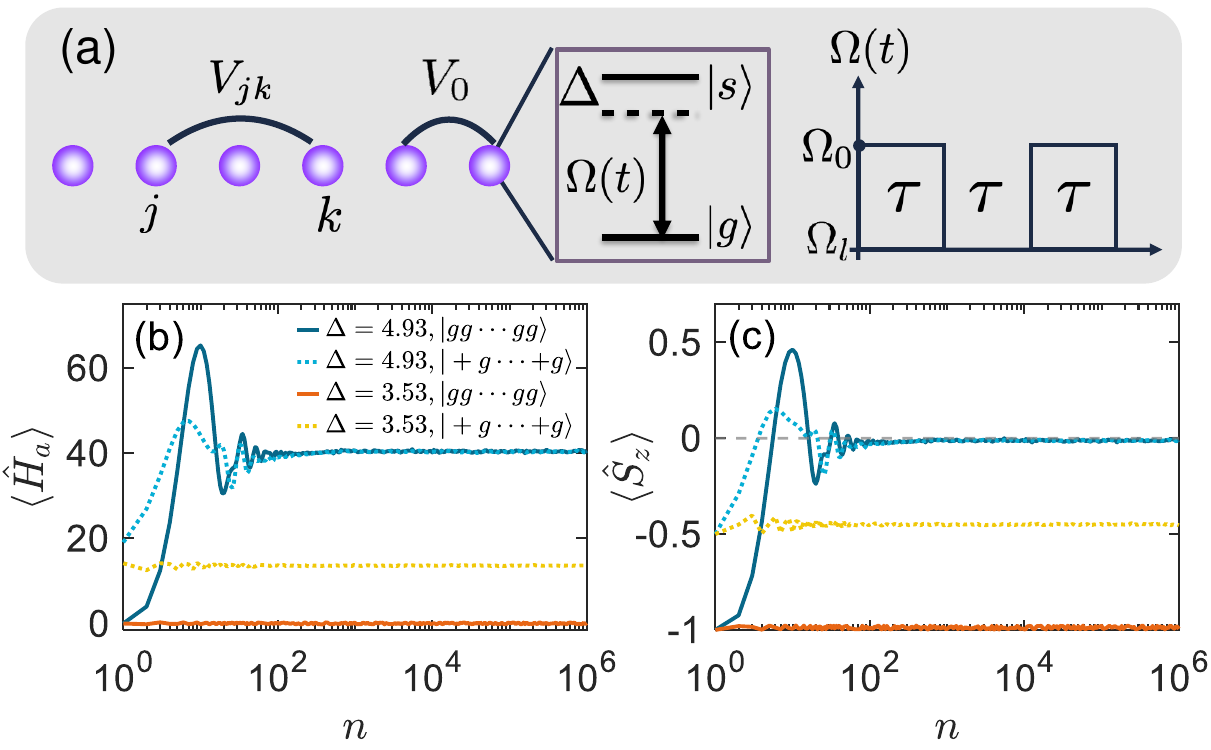}
	\caption{(a) Floquet driven Rydberg atom array. Atoms are coupled from the ground state $|g\rangle$ to Rydberg state $|s\rangle$ by a laser with detuning $\Delta$ and time-dependent Rabi frequency $\Omega(t)$. The latter forms a pulse train between $\Omega_l$ and $\Omega_0$ with a period $T=2\tau$. Rydberg atoms interact through the vdW interaction $V_{jk}$. By controlling detuning $\Delta$ and NN interaction $V_0$, the Floquet dynamics can be altered between the MBQC and integrable phases. When $(\Delta+V_0)T\approx 2K\pi$ ($K$ is an integer), many-body dynamics thermalize rapidly, as shown by stroboscopic evolution of (b) mean energy $\langle\hat{H}_a\rangle$  and (c) population $\langle\hat{S}_z\rangle$. Here $\langle \hat{H}_a\rangle\approx\epsilon_N/2$, and $\langle \hat{S}_z\rangle\approx 0$ in the thermalization phase (after about $n>100$ Floquet period). In the integrable phase, the energy and population are largely unchanged from the initial state. Parameters are $N=14, \Omega_0=1, V_0=2$ and  $\tau=\pi$. See text for details.}
	\label{fig.model}
\end{figure}

In this work, we propose an experimentally accessible Floquet engineering scheme to efficiently manipulate and control thermalization and localization in a Rydberg atom array. The protocol employs square-wave modulated laser fields, which periodically couple the electronic ground and Rydberg states with alternating high ($\Omega_0$) and low ($\Omega_l\ll \Omega_0$) Rabi frequencies of duration $\tau=T/2$ ($T$ is the period)~(Fig.~\ref{fig.model}(a)). Varying the laser detuning $\Delta$ induces a sequential onset of many-body quantum chaos (MBQC) in the system. The chaotic regime is realized when the nearest-neighbor (NN) van der Waals (vdW) interaction $V_0$ is commensurate with $\Omega_0$. Under an optimal condition $\Delta_0T=2K\pi$ ($\Delta_0=\Delta+V_0$ and $K$ is an integer), the Floquet eigenphases follow the circular orthogonal ensemble (COE)~\cite{dalessio2014longtimea}, reflecting the MBQC. Dynamically, this leads to a \textit{reciprocal Floquet thermalization}, triggered by energy absorption (Fig.~\ref{fig.model}(b)), where the atomic population saturates to near zero (Fig.~\ref{fig.model}(c)). Crucially, the effect arises in a disorder-free array, uses experimentally accessible time-dependent laser fields and initial states, and develops on timescales within the Rydberg lifetime. By combining Floquet engineering with precise control over Rydberg atom interactions, our study establishes a route for observing and steering Floquet thermalization and engineering effective many-body dynamics in Rydberg atom arrays~\cite{tianEngineeringFrustratedRydberg2025}.

\begin{figure*}[htp!]
\centering  
\includegraphics[width=0.96\textwidth]{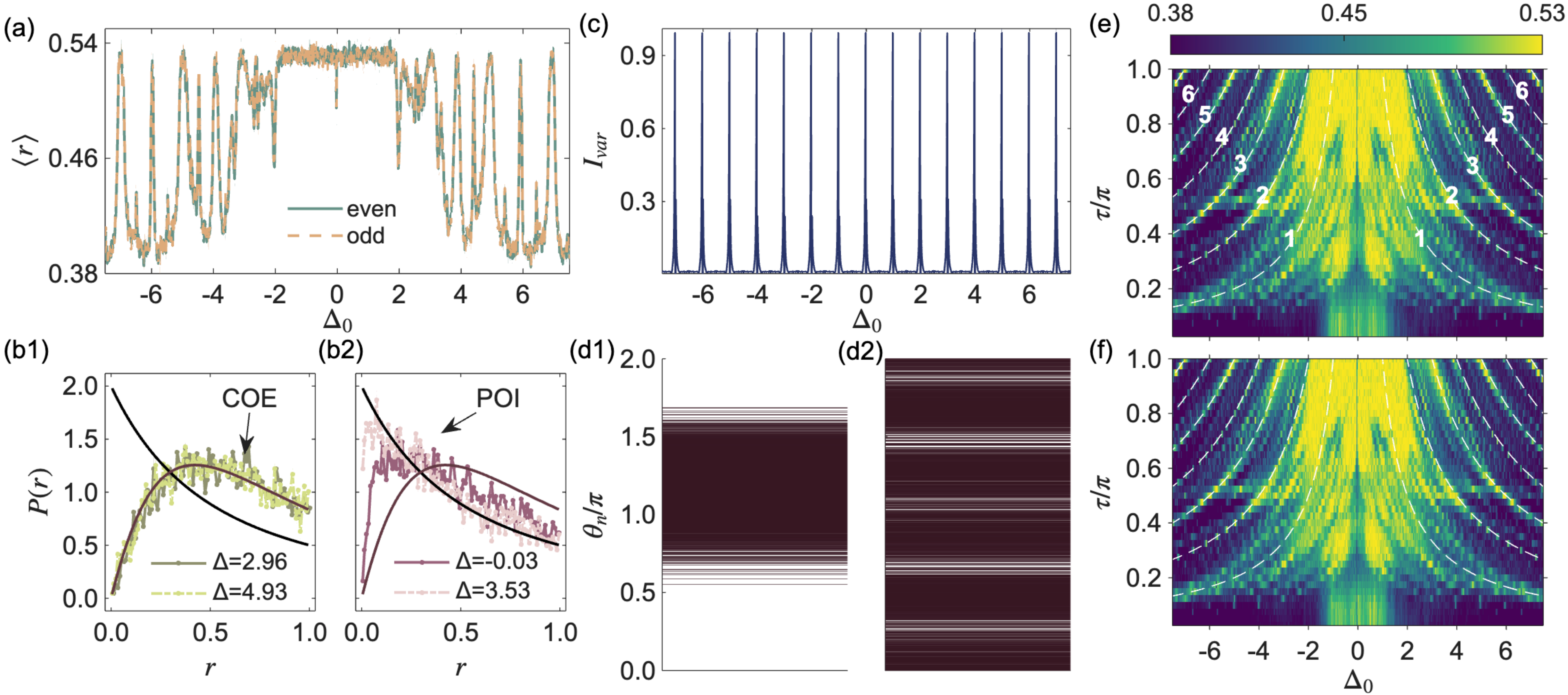}
\caption{(a) Profile of $\langle r\rangle$ as a function of $\Delta_0$. When $|\Delta_0|>2$, a series of chaos peaks, located around integer $\Delta_0$, symmetrically distribute with respect to $\Delta_0$. The even (solid) and odd (dotted) parity sectors have similar shapes and sizes. At the peaks, the level spacing exhibits COE statistics (b1), while it becomes largely a POI distribution (b2) at the lowest values (in the middle of two adjacent peaks). (c) Spreading of phase $\bar{\theta}_n$ of the undriven Hamiltonian $\hat{H}_2$. We show $I_{var}=I_0/\text{Var}(\bar{\theta}_n)$ with $\text{Var}(\bar{\theta}_n)$ to be the variance of $\bar{\theta}_n$ and $\bar{\theta}_n$ is the average value of quasi-phase $\theta_n$ over all eigenstates. We normalize $\text{Var}(\bar{\theta}_n)$ to be $I_0$ such that the maximal $I_{var}$ is 1. It can be seen that $I_{var}$ peaks when $\Delta_0$ is an integer. (d) Distribution of eigenphase $\theta_n$ of the Floquet operator. The envelope of $\theta_n$ is a Gaussian (d1) at the chaos peak, while it is banded (d2) at the valley of $\langle r\rangle$. Diagrams of $\langle r\rangle$ as a function of $\Delta_0$ and $\tau$ for even (e) and odd (f) sectors. The dashed curves, $\Delta_0T=2K\pi$, highlight the locations of the peaks. $|K|$ denotes the order of the reciprocal Floquet thermalization. In panels (a)–(d), $N=14$ and $\tau = \pi$. In (e)-(f) $N=12$. In all panels $V_0 = 2$.}
\label{fig:fig2}
\end{figure*}

\emph{Floquet driven Rydberg atom array setting---}
Our system is a one-dimensional array of $N$ atoms with open boundary conditions, as depicted in Fig.~\ref{fig.model}(a). Atoms in ground state $\ket{g}$ are excited to Rydberg state $\ket{nS}$ (with principal quantum number $n$ and orbital angular momentum $l=0$), which we denote as $\ket{s}$ for simplicity. The excitation is driven by a periodic Rabi frequency $\Omega(t)$. The Hamiltonian of the many-body system reads ($\hbar\equiv 1$)
\begin{equation}\label{eq:Hamiltonian}
\hat{H}=\frac{\Omega(t)}{2}\sum_j^N\hat{\sigma}_j^x+\Delta\sum_j^N  \hat{n}_j+\sum_{j<k}^{N}V_{jk}\hat{n}_j\hat{n}_{k},
\end{equation}
where $\Delta$ is the detuning, and operator $\hat{\sigma}_j^x=\ket{g_j}\bra{s_j}+\ket{s_j}\bra{g_j}$ and $\hat{n}_j=\ket{s_j}\bra{s_j}$ are the Pauli matrix and projection operator of the Rydberg state, respectively. Atoms in the Rydberg state interact via the vdW interaction, $V_{jk}=\frac{V_0}{|j-k|^6}$, where $V_0=C_6/a^6$ is the NN interaction ($C_6 \propto n^{11}$ and $a$ to be the dispersion coefficient and lattice spacing). In what follows, we scale the Hamiltonian and time by $\Omega_0$ and $1/\Omega_0$ except when discussing the experimental realization. 

We employ a Floquet driving scheme where the Rabi frequency $\Omega(t)$ is shaped to be a square-wave with strengths $\Omega_0$ and $\Omega_l$, and period $T=2\tau$ [see Fig.~\ref{fig.model}(a)]. While we focus on $\Omega_l=0$, our study is valid for any $|\Omega_0|\gg |\Omega_l|$, see Supplemental Material (\textbf{SM})~\cite{SM} for discussions (also including Refs.~\cite{mullenbach2025quantum,xu2020probing,britton2012engineered,richerme2014nonlocal,bohnet2016quantum,dukeszInterplayInteractionUncorrelated2009,misguichInverseParticipationRatios2016,haegeman2011timedependent,fishman2022itensor,aditya2024subspacerestricted,santos2012chaos,robertson2021arc,kuwaharaFloquetMagnusTheory2016,goldman2014periodically,jordan1928uber,joel2013introduction,howell2019asymptotic}). In Ref.~\cite{koyluoglu2025floquet}, it has been shown that the square-wave modulation of laser detuning offers control over Rydberg atom interactions and entanglement dynamics. In our ``bang-bang" protocol for the Rabi frequency, we achieve flexible control over the many-body phases and reveal thermalization-localization dynamics. Here Hamiltonian~(\ref{eq:Hamiltonian}) alternates between $\hat{H}_1 = \Omega_0/{2}\sum_j\hat{\sigma}_j^x +\hat{H}_2$ and $\hat{H}_2=\Delta\sum_j \hat{n}_j+\sum_{j<k}V_{jk}\hat{n}_j\hat{n}_{k}$. Dynamical evolution of the atoms is generated by the Floquet operator~\cite{blanes2009magnus,goldman2014periodically,ponte2015periodically},
\begin{equation}\label{floq}
\hat{U}_F=e^{-i\hat{H}_2\tau}e^{-i\hat{H}_1\tau}=e^{-i\hat{H}_{F}T},
\end{equation}
where $\hat{H}_{F}$ is the Floquet Hamiltonian~\cite{kuwaharaFloquetMagnusTheory2016}. In the high frequency regime $\tau \ll 1$, we obtain an effective quantum spin model from the leading terms of the Magnus expansion (\textbf{SM}~\cite{SM}). This model features Dzyaloshinsky-Moriya-type interactions, which could induce chiral and magnetic phases~\cite{brockmannExactDescriptionMagnetoelectric2013,radhakrishnanQuantumCoherencePlanar2017,jinQuantumPhaseTransitions2025}. We will go beyond the high-frequency limit to regimes where finite-order expansions are generally insufficient to capture the physics. 

\emph{Emergence of many-body quantum chaos---}
To understand dynamics of the Floquet driven Rydberg atom chain, we first diagonalize the Floquet operator to obtain its eigenphase $\theta_n$ and eigenstate $|\theta_n\rangle$, $\hat{U}_F\ket{\theta_n}=e^{-i\theta_n}\ket{\theta_n}\, (n=1,\cdots\mathcal{D})$, where $\mathcal{D}$ is the dimension of Hamiltonian $\hat{H}$~\cite{eckardt2015highfrequency}. The eigenphases, $\theta_{\mathcal{D}}\geq \theta_{\mathcal{D}-1} \geq\cdot\cdot\cdot \geq \theta_1$, are unfolded in the interval  $[0, 2\pi)$ and related to quasienergies $\varepsilon_n$ through $\theta_n=\varepsilon_n T$. The level spacing ratio, $r_n=[\min({\delta_n,\delta_{n+1}}]/[\max({\delta_n,\delta_{n+1}})]$ between two consecutive eigenphases $\delta_n=\theta_{n+1}-\theta_{n}$, captures signatures of quantum chaos~\cite{oganesyan2007localization,atas2013distribution,dalessio2014longtimea}. For $N\to\infty$, its average converges to $\langle r\rangle\approx 0.527$ (chaotic) when distribution of $\theta_n$ is described by COE~\cite{mehta2004random}, and to $\langle r\rangle\approx 0.386$ (integrable) with Poisson (POI) distribution~\cite{pal2010manybody,khemani2016phase}.

Due to the flexible control over laser and Rydberg interactions, we focus on an optimal situation where $V_0/\Omega_0=K$, i.e., the NN interaction is $K$ times the Rabi frequency $\Omega_0$ ($K$ to be an integer). This is important for controlling the integrable-chaos property of the Floquet-driven atom system~\cite{SM}. As an example, the average ratio $\langle r\rangle$ is shown in Fig.~\ref{fig:fig2}(a) for parameters $\tau=\pi$ and $V_0=2$. Here $\langle r\rangle$ distributes (almost) symmetrically with respect to a shifted detuning $\Delta_0=\Delta + V_0=0$. There is a broad region, $-2<\Delta_0<2$, where $\langle r\rangle$ is approximately $0.527$, i.e. the eigenphase shows COE statistics. By increasing $|\Delta_0|$, we observe a series of narrow peaks $\langle r\rangle\approx 0.527$ when $\Delta_0$ is near an integer, corresponding to the COE (Fig.~\ref{fig:fig2}(b1)). Between two neighboring peaks, $\langle r\rangle$ is low and close to that of the POI distribution, as illustrated in Fig.~\ref{fig:fig2}(b2). For the open-ended chain, the profile of the level spacing statistics is almost identical for both the even and odd parity sectors (Fig.~\ref{fig:fig2}(a))~\cite{SM}. 

To understand the mechanism of quantum chaos at these peaks, we note that the undriven Hamiltonian (e.g., set $\Omega_0=0$ in $\hat{H}$) is fully given by the classical Hamiltonian $\hat{H}_2$. Its diagonal matrix elements are approximately integers owing to the optimal parameter. In other words, the unperturbed energies of many-body basis states $|j\rangle$ are approximately integers. For example, considering basis state $|\phi_N\rangle=|ss\cdots ss\rangle$ (all atoms in state $\ket{s}$), we obtain its expectation value,  $\epsilon_N=\langle \phi_N|\hat{H}_2|\phi_N\rangle=N[\Delta + V_0L_6(N-1)-V_0L_5(N-1)/N]$ where $L_n(N)=\sum_{j=1}^N j^{-n}$ is the so-called Harmonic number of order $n$. For $N\gg 1$, $L_5(N-1)/N\to 0$ and $1<L_6(N)<L_6(\infty) = \pi^6/945\approx 1.017$. Here, the small decimal part in the latter results from the tails of the vdW interactions. Neglecting these contributions, we find that $\epsilon_N /N\approx \Delta_0$, which becomes an integer if $\Delta_0$ is an integer. The diagonal elements of $\hat{H}_2$ are around integers independent of $N$. We emphasize that the vdW interaction tails only slightly broaden the distributions of the unperturbed energies from the integers (Fig.~\ref{fig:fig2}(c)). 

In the full Floquet operator, the laser  ($\propto\hat{\sigma}_j^x$) couples the many-body basis states, generating a Gaussian envelope of $\theta_n$ (Fig.~\ref{fig:fig2}(d1)). This is a key feature of chaotic systems~\cite{haake2010quantum}. As shown in Figs.~\ref{fig:fig2}(e) and (f), chaos peaks are found when $\Delta_0 T=2K\pi$. This results to a series of \textit{reciprocal quantum chaos} in the vicinity of $\Delta_0 T=2K\pi$, in contrast to many previous studies where celebrated smooth integrable-to-chaos transitions are found~\cite{oganesyan2007localization,atas2013distribution,dalessio2014longtimea,zhang2016floquet,sonner2021thouless}. Away from the chaos peak, the eigenvalues of $\hat{H}_2$ have broad distributions (Fig.~\ref{fig:fig2}(c)). When the Floquet coupling is switched on, the distributions of eigenphase $\theta_n$ turn out to be banded (Fig.~\ref{fig:fig2}(d2)). Different bands form local subsystems, such that quantum chaos is prohibited~\cite{lazaridesFateManyBodyLocalization2015}.

We briefly make two remarks before proceeding to the next section. First, the qualitative difference between the regimes $|\Delta_0|<2$ and $|\Delta_0|>2$ originates from the way the drive couples to the banded quasienergy spectrum of $\hat{H}_2$. For small detuning, the drive strongly hybridizes multiple Floquet sidebands, effectively washing out the band structure and enabling rapid energy redistribution, which is indicative of quantum chaos. In contrast, when $|\Delta_0|$ exceeds this threshold, the band spacing becomes comparable to or larger than the Rabi coupling $\Omega_0$, such that the drive predominantly couples near-resonant subbands. Second, a few lower peaks appear between two adjacent peaks corresponding to the reciprocal quantum chaos (for example, the one between $\Delta_0\approx4$ and $5$ in Fig.~\ref{fig:fig2}(a)). This is mainly attributed to the tail of the vdW interaction~\cite{SM}.

\emph{Reciprocal Floquet thermalization---} We explore thermalization and localization by studying the stroboscopic dynamics of the Rydberg atom array. After $n$ periods, expectation value $\langle\hat{O}\rangle(nT) = \langle \psi(nT)|\hat{O}|\psi(nT)\rangle$ of operator $\hat{O}$ is evaluated with stroboscopic state $|\psi(nT)\rangle = \hat{U}(nT)|\psi(0)\rangle$, where evolution operator $\hat{U}(nT)=[\hat{U}_F(T)]^n$. Normally, initial states typically lie in the middle region of the eigenspectrum~\cite{torres-herreraEffectsInterplayInitial2013,heInitialstateDependenceQuench2013} to ensure thermalization. These initial states are typically difficult to prepare experimentally. We employ two initial states $|\phi_0\rangle=|gg\cdots gg\rangle$, i.e. all atoms are state $\ket{g}$, and $|\phi_1\rangle=|+g\cdots+g\rangle$ with $|+\rangle = (|g\rangle+|s\rangle)/\sqrt{2}$, i.e. atoms are in superposition $|+\rangle$ and $\ket{g}$ alternatively in the array. These product states can be prepared experimentally~\cite{shen2023proposal,zhao2025observation}.  

\begin{figure}[htp!]
	\centering  
	\includegraphics[width=1\linewidth]{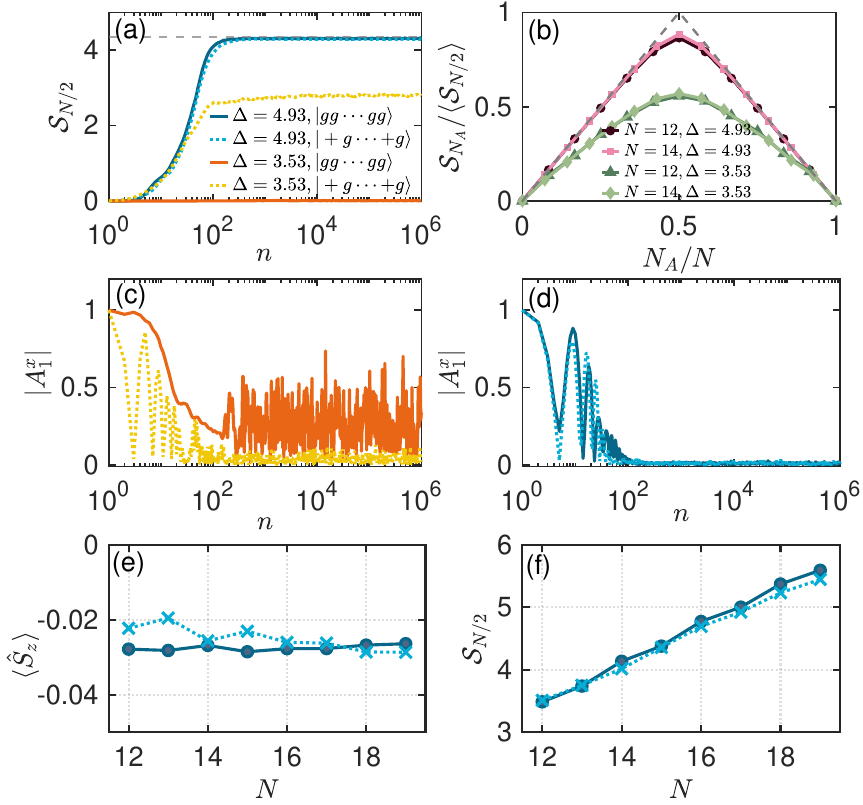}
	\caption{(a) Stroboscopic  dynamics of the half-chain entropy $\mathcal{S}_{N/2}$, which saturates to the Page value due to thermalization. (b) Saturated entanglement entropy of subsystem $A$. In the chaotic case $\Delta=4.93$, the entropy follows the volume law and saturates to the Page value. In the integrable case $\Delta=3.53$, the entropy is lower than the Page value. The entropies for $N=12$ and 14 are almost identical in the corresponding phases. The entropy is averaged in the range $n \in [10^3,10^4]$. Dynamics of $|A_1^x|$ in the integrable (c) and chaotic phase (d). The edge mode fluctuates strongly and persistently over time in (c). It decays quickly when the dynamics are thermalized. Saturated population $\langle \hat{S}_z\rangle$ (e) and half-chain entropy $\mathcal{S}_{N/2}$ (f) obtained using TDVP calculations for chaotic case ($\Delta=4.93$). In (a), (c) and (d), we consider $N=14$. Other parameters are $V_0=2$ and $\tau=\pi$. }
	\label{fig:fig3}
\end{figure}

We first show average population $\langle\hat{S}_z\rangle(nT)=\bra{\psi(nT)}1/N\sum_j\hat{\sigma}_j^z\ket{\psi(nT)}$ in Fig.~\ref{fig.model}(c). In the chaotic regime, the population grows from $-1$ ($-0.5$) using the initial state $|\phi_0\rangle$ ($|\phi_1\rangle$). It reaches a steady value, around $0$, as expected in a thermalizing state~\cite{bukov2016heating,ren2020noisedriven}. At the same time, the entanglement entropy saturates at the maximal possible value. We calculate the von Neumann entropy of subsystem $A$ at stroboscopic time $nT$, $\mathcal{S}_{N_A}(nT)=-\text{Tr}_{A}~[\rho_A(nT)\ln\rho_A(nT)]$, where $N_A$ ($N_B=N-N_A$) denotes the left (right) sites, and $\rho_A(nT)=\text{Tr}_{N_B}(|\psi(nT)\rangle\langle \psi(nT)|)$ is the reduced density matrix of subsystem $A$. The half-chain entropy $\mathcal{S}_{N/2}$ saturates to the Page value $\langle\mathcal{S}_{N/2}\rangle=\ln{2^{N/2}}-1/2$~\cite{page1993average}, as depicted in Fig.~\ref{fig:fig3}(a). Varying $N_A$, the saturated entropy increases linearly with $N_A<N/2$ (Fig.~\ref{fig:fig3}(b)) as a result of the volume law in thermal states~\cite{deutsch2018eigenstate}.

In the integrable phase, the energy and population are confined to the initial values (Fig.~\ref{fig.model}(b)-(c)). The entanglement entropy is much lower than that of thermalization (Fig.~\ref{fig:fig3}(a) and (b)). Despite the localization, strong fluctuations are found in edge modes, characterized by two-time correlation $|A_1^x|=\bra{\psi(0)}\hat{\sigma}_1^x(nT)\hat{\sigma}_1^x\ket{\psi(0)}$~\cite{yates2019almost}, as depicted in Fig.~\ref{fig:fig3}(c). In contrast, the edge mode decays to zero in the thermalizing phase (Fig.~\ref{fig:fig3}(d)). 

Other quantities, such as auto-correlation~\cite{aditya2024subspacerestricted}, inverse participation ratio~\cite{dukeszInterplayInteractionUncorrelated2009}, and principal components~\cite{santos2012chaos}, lead to consistent expectations with the population and entropy (see \textbf{SM}~\cite{SM}). Moreover, we perform a finite-size scaling analysis in the thermal phase. In Figs.~\ref{fig:fig3}(e) and (f), we show saturated $\langle \hat{S}_z\rangle$ and $\mathcal{S}_{N/2}$ obtained using time-dependent variational principle (TDVP) calculations~\cite{haegeman2011timedependent,fishman2022itensor}. The former is approximately $-0.02$, while the latter increases with $N$ in the thermal cases, indicating that the initial states cause robust thermalization dynamics~\cite{SM}.

\begin{figure}[htp!]
   \centering  
   \includegraphics[width=0.99\linewidth]{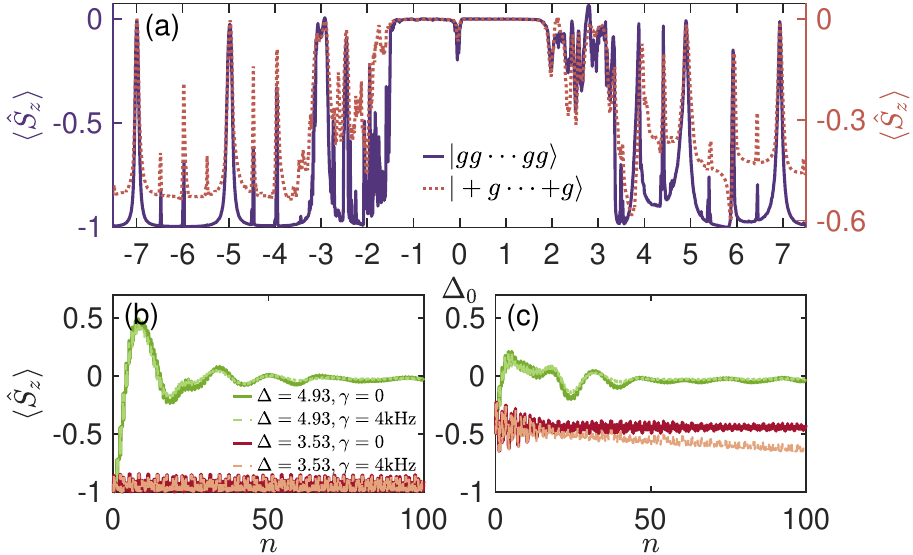}
	\caption{(a) Average population $\langle\hat{S}_z\rangle$ for initial state $|\phi_0\rangle$ (left, solid line), and $|\phi_1\rangle$ (right, dotted line). Time evolution of the average population $\langle\hat{S}_z\rangle$ for initial state $|\phi_0\rangle$ (b) and $|\phi_1\rangle$ (c). Rydberg decay (dashed) plays a negligible role during the thermalization, but is state-dependent in the integrable regime. For state $|\phi_1\rangle$, the deviation results from exponential decay of the Rydberg atoms in the otherwise frozen initial state. We consider $N=14$ in panel (a) and $N=12$ in panels (b) and (c). Other parameters are same with Fig.~\ref{fig:fig3}.}
	\label{fig:fig4}
\end{figure}

\emph{Experimental probing of the thermalization and localization---} 
As the population $\langle \hat{S}_z\rangle$ is sensitively dependent on the many-body phase, this provides a practical way to probe the reciprocal Floquet thermalization and localization dynamics. In Fig.~\ref{fig:fig4}(a), we show average population $\langle \hat{S}_z\rangle$ when the dynamics saturates. The sharp population peaks coincide with the thermalization peaks. In the integrable region, the average population is much smaller than 0. The distinctive population distribution thus encodes and provides a way to measure the level spacing statistics shown in Fig.~\ref{fig:fig2}(a).

Crucially, the dynamics of thermalization and localization remain observable within the finite lifetimes of Rydberg states. We demonstrate this using the $|60S\rangle$ state of Rb atoms, whose lifetime is $230\,\mu$s (decay rate $\gamma=4.34\,$kHz) and $C_6\approx 139~\mathrm{GHz}\cdot(\mu\mathrm{m})^6$~\cite{robertson2021arc}. With a lattice spacing $a=4.9 \,\mu$m, the resulting NN interaction is $V_0\approx10$~MHz. For $\Omega_0=5$~MHz and $\tau = 0.2 \pi\,\mu$s, this leads to $V_0/\Omega_0=2$, $\Omega_0T=2\pi$ and scaled decay rate $\gamma/\Omega_0\approx8\times10^{-4}$. We then solve the underlying master equation for the different phases~\cite{SM}. As shown in Figs.~\ref{fig:fig4}(b)-(c), the decay has a negligible effect up to $n=100$ (corresponding to $125.7~\mu$s). Importantly the saturation of the population is clearly visible. In Fig.~\ref{fig:fig4}(c), the deviation of the population in the integrable phase results from the exponential decay of the Rydberg component in the localized initial state. The consistent results provide a reliable basis for characterizing the two phases.

\emph{Conclusion and Discussion---} We have investigated a Floquet scheme that enables the control of the reciprocal Floquet thermalization and localization dynamically in a one-dimensional Rydberg atom chain.  The conditions to trigger and the mechanism of the thermalization are identified, which rely on the optimal combination of the strong vdW interaction and laser parameters. The vdW interaction is a key feature to generate the narrow chaos peaks, as other types of interactions (e.g. dipolar or Coulomb interactions) normally give broad chaos phases~\cite{SM}. This reveals that the exponent of the interaction potential is a central parameter for controlling non-equilibrium dynamics and emergent phases~\cite{mullenbach2025quantum}. We have also shown that thermalization and localization occur on experimentally accessible timescales and with the initial states used in Rydberg atom array experiments~\cite{SM}. Future investigations using sophisticated approaches~\cite{regnault2025integer,schindlerGeometricFloquetTheory2025} may reveal the geometric origin of the reciprocal Floquet thermalization. The ability to control dynamical localization and thermalization has potential applications in developing digital quantum simulation~\cite{heylQuantumLocalizationBounds2019,sieberer2019digital} with Rydberg atoms~\cite{saffman2010quantum, weimerDigitalQuantumSimulation2011,bohrdtMultiparticleInteractionsUltracold2020,morgadoQuantumSimulationComputing2021a,wuConciseReviewRydberg2021,grahamMultiqubitEntanglementAlgorithms2022,cesaUniversalQuantumComputation2023}. Our study furthermore opens opportunities to study edge modes~\cite{yates2019almost}, detect prethermalization~\cite{fleckenstein2021thermalization,ghosh2023detecting} and engineer effective many-body Hamiltonian using the Rydberg atom array setting~\cite{tianEngineeringFrustratedRydberg2025,koyluoglu2025floquet}. 

\emph{Acknowledgments---}
We appreciate the insightful discussions with Marin Bukov, Keiji Saito, Tianyi Yan, Li You, Huanqian Loh, Yongqiang Li, and Thomas Pohl. This work was supported by the National Natural Science Foundation of China (Grants No. U2341211, No. 62175136, No. 12241408, and No. 12120101004); Innovation Program for Quantum Science and Technology (Grants No. 2023ZD0300902); Fundamental Research Program of Shanxi Province (Grants No. 202303021224007); the 1331 project of Shanxi Province. W.L. acknowledges financial support from the EPSRC (Grant No. EP/W015641/1) and the Going Global Partnerships Programme of the British Council (Contract No. IND/CONT/G/22-23/26), and the use of the University of Nottingham's Ada HPC service.

\emph{Data availability---}
The data that support the findings of this study are openly available~\cite{Data}.



%% file: SM.tex
\section{Energy Spectral diagnostics of chaos and integrability}
\subsection{Level spacing ratio}

\begin{figure}[htp!]
\centering  
\includegraphics[width=0.65\linewidth]{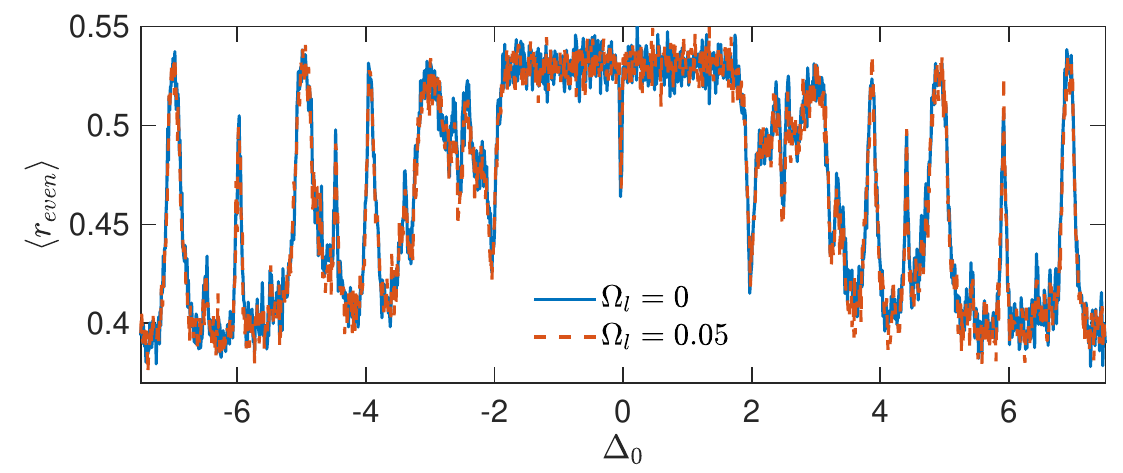}
\caption{Average level spacing ratio $\langle{r_{even}}\rangle$ as a function of $\Delta_0$ with different Rabi frequency $\Omega_l$. Other parameters are $N=12$, $V_0=2$, and $\tau=\pi$.}
\label{r_compareOmega}
\end{figure}

In our Floquet scheme, we implement periodic modulation of the laser Rabi frequency such that $\Omega(t)$ alternates between $\Omega_0$ and $\Omega_l$. In the main text, we focused on the situation $\Omega_l = 0$. We calculate the adjacent level spacing ratio of the system, as shown in Fig. 2(a) of the main text. Our exact diagonalization (ED) calculation is valid even when $\Omega_l\neq 0$ as long as $\Omega_0 \gg \Omega_l$. As illustrated in Fig.~\ref{r_compareOmega}, we compute the adjacent level spacing ratio $\langle{r_{even}}\rangle$ as a function of $\Delta_0$ for $\Omega_l = 0.05$. The profile is nearly identical to that of the $\Omega_l = 0$ case.

\begin{figure}[htp!]
\centering  
\includegraphics[width=0.99\linewidth]{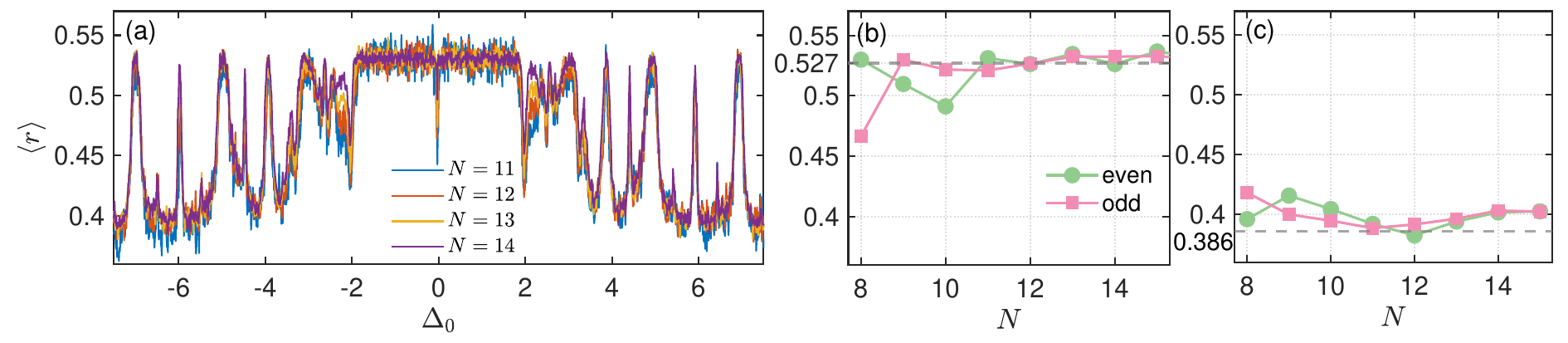}
\caption{(a) Average level spacing ratio $\langle{r_{even}}\rangle$ as a function of $\Delta_0$ with different system scale. The finite-size effect for (b) chaotic phase $\Delta=4.93$ and (c) integrable phase $\Delta=3.53$. The gray dashed lines represent the predicted value for COE and POI distribution. Other parameters are $V_0=2$ and $\tau=\pi$.}
\label{r_compareN}
\end{figure}

In the main text, we calculated the average level spacing ratio $\langle{r_{even}}\rangle$ with $N = 12$ and $N = 14$. Reciprocal Floquet thermalization is not exclusive to systems with even $N$. To demonstrate this, we compute the adjacent level spacing ratio for odd particle numbers $N = 11$ and $N = 13$, and compare it with the cases of $N = 12$ and $N = 14$, shown in Fig.~\ref{r_compareN}(a). The ED calculation shows that the positions of the peaks remain unchanged for different particle numbers. Furthermore, we investigate the finite-size effects in two distinct phases, as shown in Fig.~\ref{r_compareN}(b) and (c), respectively. When $\Delta=4.93$, the system is in the chaotic phase, and the average level spacing ratio $\langle{r}\rangle$ oscillates around $0.527$ as the number of atoms increases. In contrast, for $\Delta=3.53$, although $\langle{r}\rangle$ is slightly greater than $0.386$, it remains significantly lower than the COE prediction. This stable deviation from the random matrix theory prediction provides strong evidence that the non-thermalizing behavior at this point is not a finite-size artifact, but a robust manifestation of an underlying non-ergodic phase in the thermodynamic limit.

\begin{figure}[htp!]
\centering  
\includegraphics[width=0.65\linewidth]{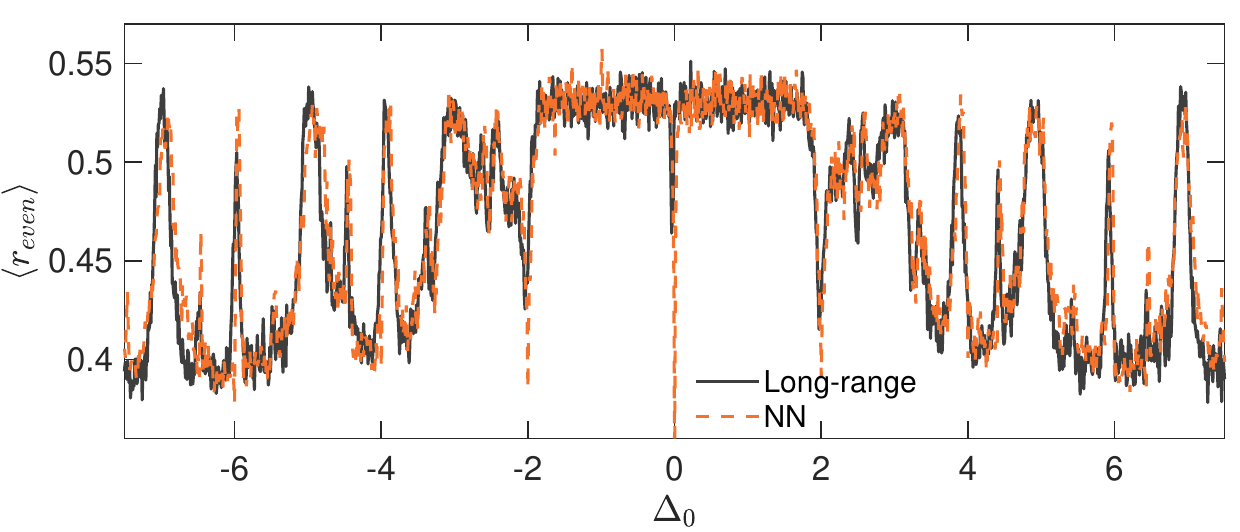}
\caption{Average Level spacing ratio $\langle{r_{even}}\rangle$ as a function of $\Delta_0$ with different interaction type. Other parameters are $N=12$, $V_0=2$ and $\tau=\pi$.}
\label{Vrange}
\end{figure}

We have considered a chain of Rydberg atoms, where our model takes into account the long-range interactions between Rydberg atoms. However, as the vdW interaction decays rapidly with distance, the interactions between nearest-neighbor (NN) atoms are important. As shown in Fig.~\ref{Vrange}, 
the tail of the vdW interaction only marginally modifies the level spacing statistics.  When $|\Delta_0|>3$, the tail slightly modifies the height and width of the peaks. When only NN interactions are included, the level spacing ratio reaches its minimum at $\Delta_0 = 0$, where the system is integrable (see the discussion on the effective Hamiltonian below).

\begin{figure}[htp!]
\centering  
\includegraphics[width=0.6\linewidth]{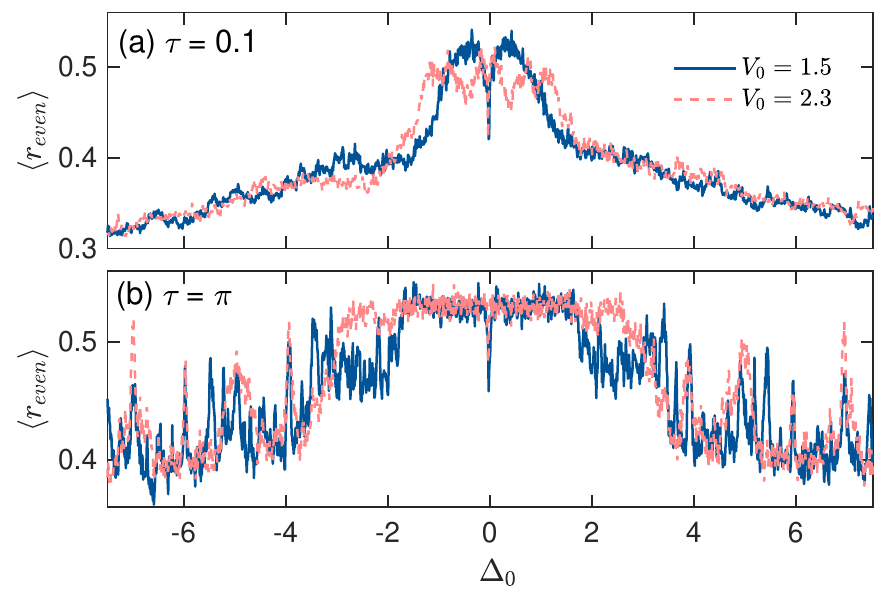}
\caption{Average level spacing ratio $\langle{r_{even}}\rangle$ as a function of $\Delta_0$ with different NN interaction strength $V_0$ for pulse duration (a) $\tau=0.1$ and (b) $\tau=\pi$ when $N=12$.}
\label{V_strength}
\end{figure}

The reciprocal Floquet thermalization depends sensitively on the optimal condition, $V_0 = K'\Omega_0$ with $K'$ being an integer. To show this, we examined the level spacing statistics when $K'$ is not an integer. As shown in Fig.~\ref{V_strength}, we examine two specific cases, $K' = 1.5$ and $K' = 2.3$. The average level spacing ratio for the two different pulse durations is shown. For $\tau = 0.1$, quantum chaos is found within the range $-2 < \Delta_0 < 2$, whereas outside this region, the system tends toward integrability. Notably, no chaotic resonance peaks are observed. When $\tau = \pi$, some peaks are found, whose locations depend strongly on $V_0$.  

\begin{figure}[htp!]
	\centering  
	\includegraphics[width=0.65\linewidth]{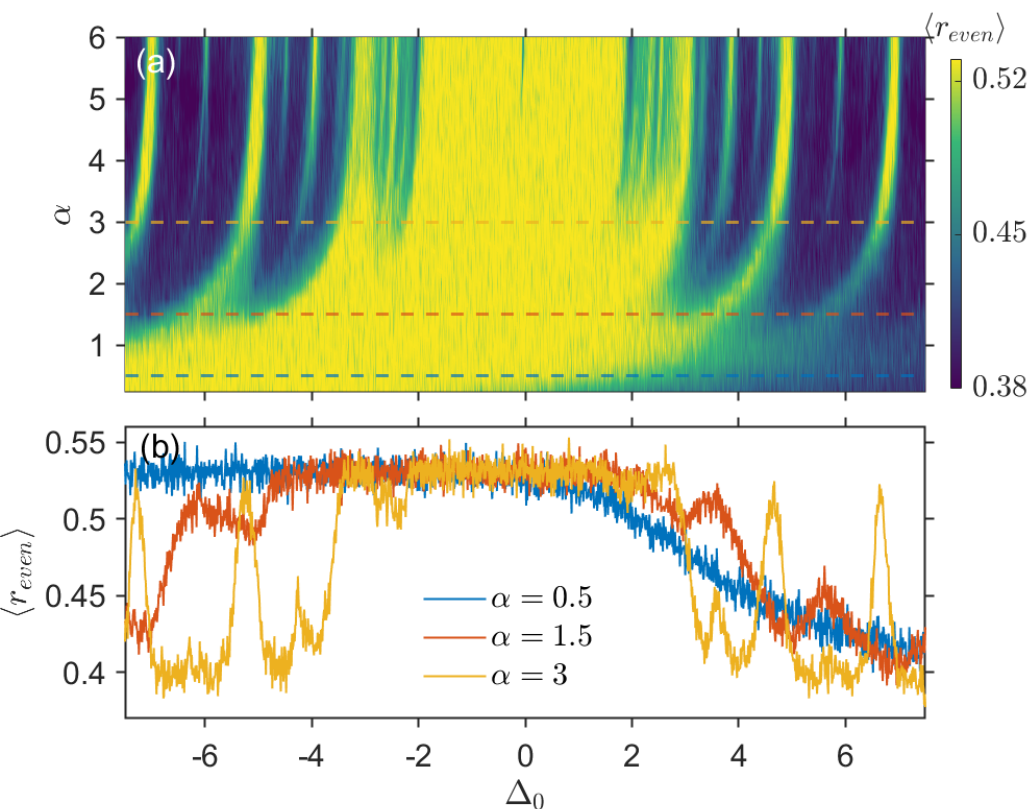}
	\caption{Dependence of thermalization signatures on the interaction exponent $\alpha$. (a) Average level spacing ratio $\langle{r_{even}}\rangle$ as a function of $\Delta_0$ and interaction coefficient $\alpha$ ($V_{jk}=V_0/R_{jk}^\alpha$). (b) Cross-sections of $\langle r_{{even}}\rangle$ for $\alpha=0.5, 1.5,$ and $3$. Exponents typical of other platforms ($\alpha \leqslant 3$) show broad, featureless behavior, confirming that the specific strong vdW interaction ($\alpha=6$) is crucial for the unique thermalization dynamics observed. Other parameters are $N=12$, $V_0=2$ and $\tau=\pi$.}
	\label{Fig:different alpha}
\end{figure}
The exponent $\alpha$ of the interaction potential has been identified as a central parameter controlling non-equilibrium dynamics and emergent phases~\cite{mullenbach2025quantum}. To analyze the impact of different interaction types on energy spectrum, we examine the average level spacing ratio across a range of parameters. Fig.~\ref{Fig:different alpha}(a) shows the colormap of the level spacing ratio $\langle r_{{even}} \rangle$ as a function of  $\Delta_0$ and $\alpha$, where $V_{jk}=V_0/R_{jk}^\alpha$. As $\alpha$ increases, pronounced reciprocity thermalization peaks and a clear spectral symmetry emerge. These features highlight the interaction characteristics that are unique to the Rydberg atom platform. This distinctive feature creates a situation qualitatively different from that in superconducting circuits, where all-to-all couplings~\cite{xu2020probing} ($V_{jk}=V_0$) induce pronounced degeneracies, and from trapped-ion systems, where interactions typically follow $V_{jk}\propto1/R_{jk}^{\alpha}$ with $0\leqslant\alpha\leqslant 1.5$ due to experimental constraints like limited laser power and decoherence~\cite{britton2012engineered,richerme2014nonlocal,bohnet2016quantum}. In Fig.~\ref{Fig:different alpha}(b), we demonstrate three different interaction types $\alpha=0.5, 1.5,$ and $3$. When $\alpha \leqslant  1.5$, the spectrum exhibits a broad overall envelope without a distinct thermalization peak. In contrast, for $\alpha = 3$, corresponding to the dipole-dipole interaction, a thermalization peak similar to that in our model appears. However, this peak is less pronounced than that in our results, and there is no visible dip near $\Delta_0 = 0$ that would indicate a tendency toward integrability. This robust comparison confirms that the specific vdW interaction is essential for the unique phenomenon reported herein and further demonstrates the uniqueness of our computational results for the Rydberg system.

\subsection{Inverse participation ratio}

To further substantiate the spectral signatures of chaos and thermalization, we evaluate the inverse participation ratio (IPR)
\begin{equation}
T_I = \sum_{k,n} \left| \langle k | \theta_n \rangle \right|^4 ,
\end{equation}
which quantifies the degree of localization of the Floquet eigenstates $|\theta_n\rangle$ in a given computational basis $\{|k\rangle\}$~\cite{dukeszInterplayInteractionUncorrelated2009}. The IPR characterizes how extensively an eigenstate spreads over the available basis states: for a fully delocalized (chaotic) eigenstate, $T_I/\mathcal{D}\to 0$, whereas for a localized eigenstate typical of integrable dynamics, $T_I/\mathcal{D}$ remains of order unity.

\begin{figure}[htp!]
\centering  
\includegraphics[width=0.65\linewidth]{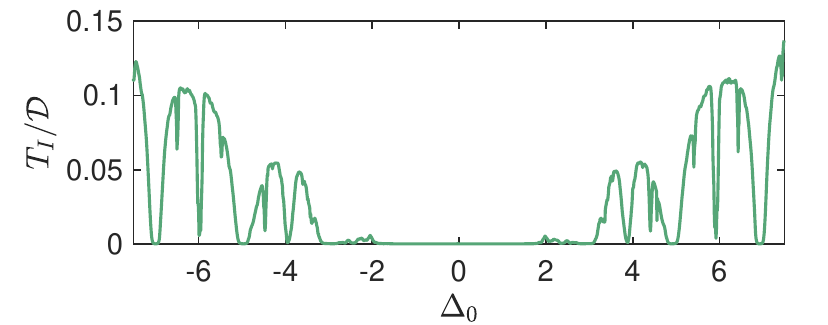}
\caption{Inverse participation ratio $T_I/\mathcal{D}$ as a function of $\Delta_0$. Other parameters are $N=14$, $V_0=2$ and $\tau=\pi$.}
\label{Fig:IPR}
\end{figure}

Figure~\ref{Fig:IPR} shows the inverse participation ratio $T_I/\mathcal{D}$ as a function of the parameter $\Delta_0$. In the parameter regime identified as chaotic by the level spacing ratio, $T_I/\mathcal{D}$ is strongly suppressed, indicating that the Floquet eigenstates are highly extended in the Hilbert space~\cite{misguichInverseParticipationRatios2016}. Remarkably, the pronounced dips in $T_I/\mathcal{D}$ coincide with the peaks in the average level spacing ratio $\langle r_{even}\rangle$ [see Fig.~2(a)] and average population $\langle \hat{S}_z \rangle$ [see Fig.~4(a) in the main text], providing independent evidence that the observed relaxation behavior originates from the delocalization of Floquet eigenstates and the resulting thermalization of the system.

In contrast, outside the chaotic regime, the IPR remains comparatively large, reflecting the persistence of localized eigenstate structures and the suppression of thermalization, consistent with the integrable behavior inferred from spectral statistics.

\section{Time evolution behaviors}

\begin{figure}[htp!]
\centering  
\includegraphics[width=0.99\linewidth]{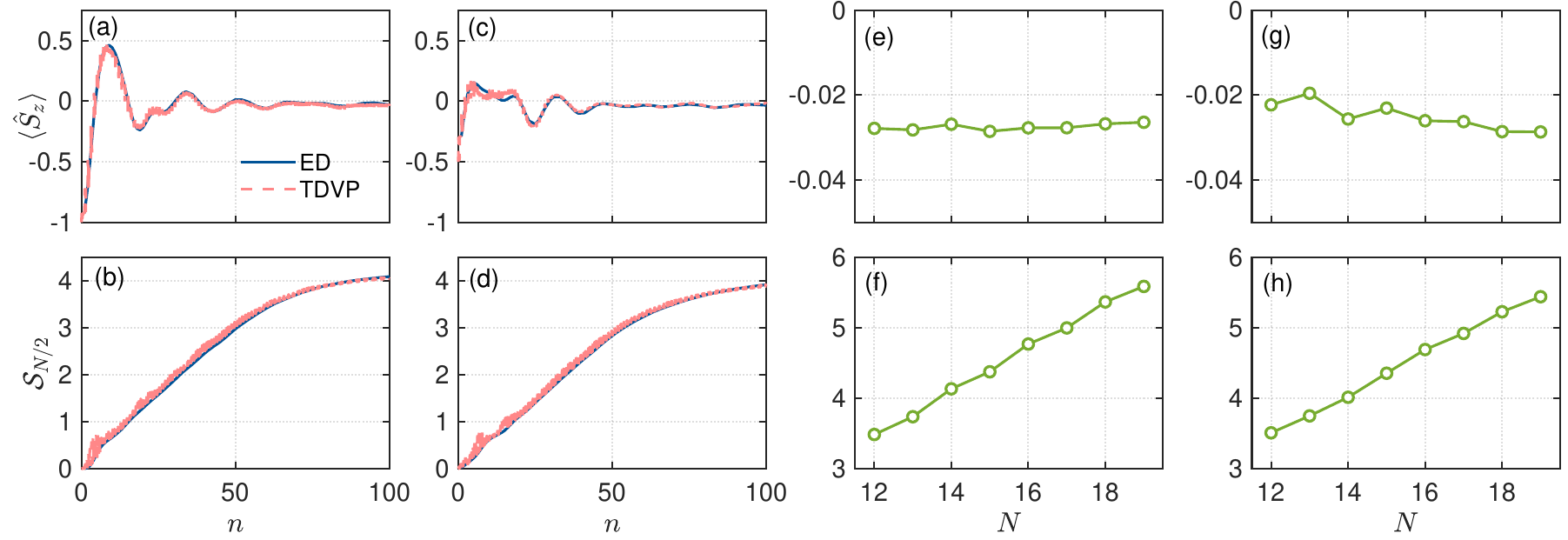}
\caption{Comparison between ED and TDVP for $N=14$ and system-size scaling of the thermalizing dynamics. The time evolution of the population $\langle {\hat{S}_z}\rangle$ (top) and the half-chain entropy $\mathcal{S}_{N/2}$ (bottom) for the initial states $\ket{\phi_0}$ (a,b) and $\ket{\phi_1}$ (c,d) demonstrate excellent agreement between the ED and TDVP methods. (e–h) Display the time-averaged values over $n\in [100,150]$ for system sizes $N=12-19$, with $\langle {\hat{S}_z}\rangle$ remaining near zero and $\mathcal{S}_{N/2}$ increasing with $N$. Other parameters are $\Delta=4.93$, $V_0=2$ and $\tau=\pi$. }
\label{MPS}
\end{figure}

To explore the thermalizing regime beyond the system sizes accessible to ED, we benchmark the time-dependent variational principle (TDVP) at $N=14$, where both methods can be applied~\cite{haegeman2011timedependent,fishman2022itensor}. Figs.~\ref{MPS}(a–d) compare the dynamical behavior of the average population $\langle\hat{S}_z\rangle$ (top panel) and the half-chain entanglement entropy $\mathcal{S}_{N/2}$ (bottom panel) for two different initial states, $|\phi_0\rangle=|gg\cdots gg\rangle$ (a-b) and $|\phi_1\rangle=|+g\cdots +g\rangle$ (c-d). In the TDVP algorithm, we employ a maximum bond dimension $\mathcal{D}_{max}=2^{N^\prime}$ with $N^\prime$ denoting the integer part of $N/2$ and a time step of $\Delta t=2\pi/50$, ensuring that at each step, the errors were below $10^{-8}$. The excellent agreement between TDVP and ED confirms the reliability of TDVP in the thermalizing phase.

We further explore the TDVP calculations for system sizes from $N=12$ to $N=19$. To characterize the late-time behavior, we compute the time-averaged population $\langle\hat{S}_z \rangle$ over the driving cycles $n\in[100,150]$, as shown in Figs.~\ref{MPS}(e-g) for $|\phi_0\rangle$ and $|\phi_1\rangle$, respectively. In both cases, $\langle\hat{S}_z \rangle$ remains close to zero and exhibits little dependence on the system size, indicating convergence to a thermal steady state. The half-chain entanglement entropy, plotted in Figs.~\ref{MPS}(f-h), grows steadily with the system size, consistent with the volume law scaling.

As a diagnosis of the presence of integrability, we investigate the behavior of the time-dependent local auto-correlation function~\cite{aditya2024subspacerestricted}
\begin{equation}    C_j(t)=\langle\psi(0)|\hat{S}_j^z(t)\hat{S}_j^z(0)|\psi(0)\rangle,
\end{equation}
where $|\psi(0)\rangle$ is the initial state of the system. This quantity measures the memory of the initial local spin configuration and serves as a sensitive probe of thermalization dynamics.

\begin{figure}[htp!]
\centering  
\includegraphics[width=0.99\linewidth]{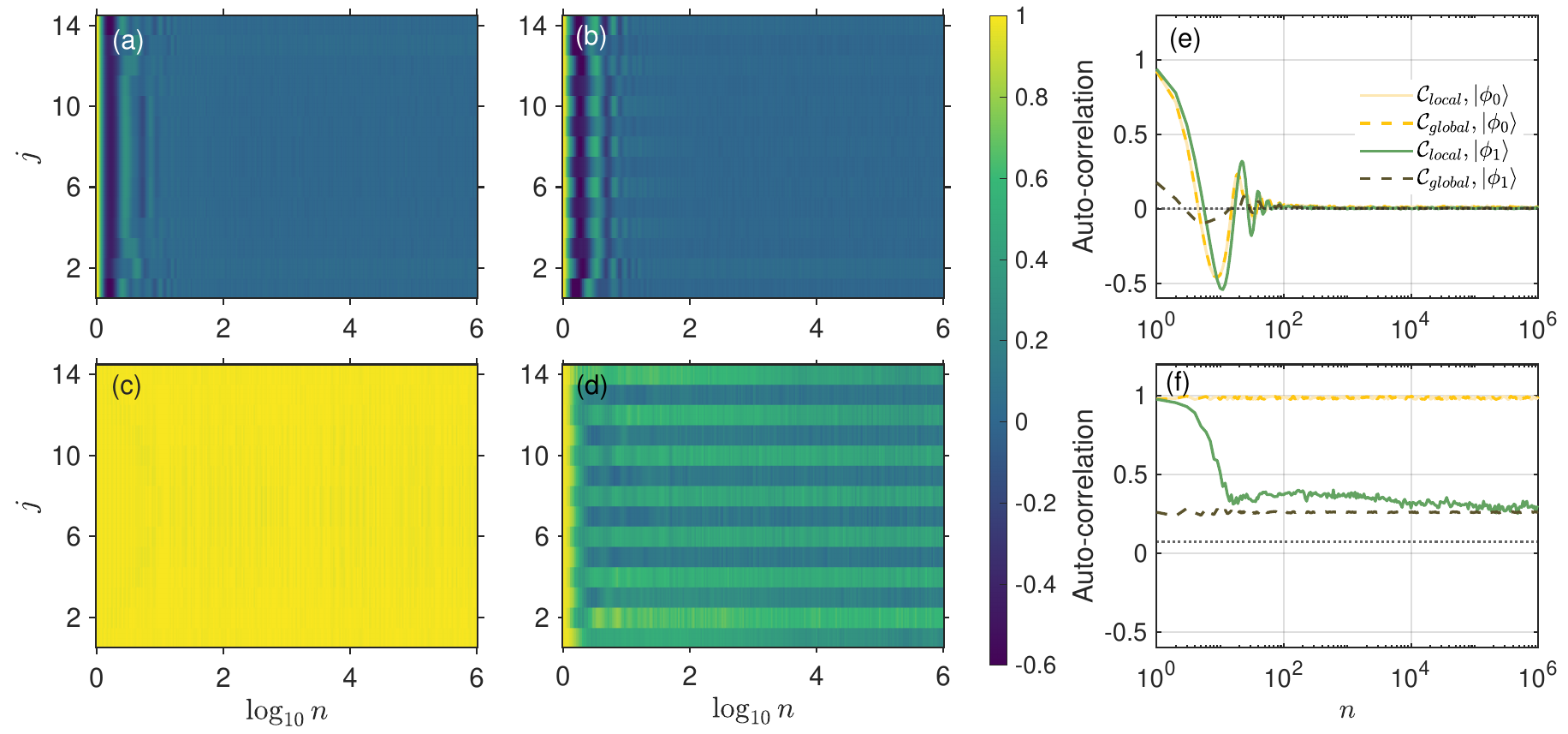}
\caption{Stroboscopic dynamics of the local auto-correlation function in the chaotic phase ($\Delta=4.93$, top panel) and the integrable phase ($\Delta=3.53$, bottom panel). (a) and (c) correspond to the initial state $|\phi_0\rangle$, while (b) and (d) correspond to the superposition of the initial state $|\phi_1\rangle$. (e) and (f) show the temporal decay of the local and global autocorrelation functions, respectively, compared with the Mazur bound (dotted lines), demonstrating clear differences between thermalizing and non-thermalizing dynamics. Other parameters are $N=14$, $V_0=2$ and $\tau=\pi$.}
\label{auto_correlator}
\end{figure}

Fig.~\ref{auto_correlator} shows the stroboscopic dynamics of the local auto-correlation function for the chaotic ($\Delta=4.93$) and integrable ($\Delta=3.53$) phases. In the chaotic regime (top panel), the local auto-correlation rapidly decays with time, indicating that the system quickly loses memory of its initial configuration and approaches a thermal steady state. Specifically, for the initial ground state $|\phi_0\rangle$ [Fig.~\ref{auto_correlator}(a)], $C_j(t)$ diminishes uniformly to nearly zero across all sites, while for the superposition initial state $|\phi_1\rangle$ [Fig.~\ref{auto_correlator}(b)], short-time oscillations appear before the correlations fully dephase at longer times. These behaviors reflect the fast relaxation and incoherent dynamics characteristic of chaotic evolution.

In contrast, the integrable regime ($\Delta=3.53$) displays persistent correlations, as shown in the bottom panel. When initialized in the state $|\phi_0\rangle$ [Fig.~\ref{auto_correlator}(c)], the system remains almost frozen with $C_j(t)\approx 1$ throughout the evolution, indicating the absence of relaxation effects. For the superposition initial state $|\phi_1\rangle$ [Fig.~\ref{auto_correlator}(d)], the local correlations exhibit long-lived oscillations that survive over a long period of time, indicating quasi-periodic motion constrained by integrability. This sustained temporal coherence demonstrates the lack of thermalization and the presence of nontrivial conserved quantities. To further quantify these observations, Figs.~\ref{auto_correlator}(e) and \ref{auto_correlator}(f) show the averaged local and global auto-correlation functions, 
\begin{equation}
    C_{local}(t)=\frac{1}{N}\sum_jC_j(t), \,C_{global}(t)=\langle \hat{S}_z(t)\hat{S}_z(0)\rangle,
\end{equation}
compared with the Mazur bound ($C_A(\infty)\equiv\frac{1}{D}\sum_a |\langle\phi_a|A|\phi_a\rangle|^2$, dotted line), which sets the lower limit imposed by conserved quantities. In the chaotic phase, the correlations decay rapidly and saturate near the bound, confirming the dominance of thermalizing processes and partial dephasing. In contrast, in the integrable regime, both $C_{local}(t)$ and $C_{global}(t)$ remain finite and well above the bound even at long times, evidencing the preservation of memory and the failure of thermalization. These contrasting behaviors provide a clear dynamical signature distinguishing the chaotic and integrable phases of the driven Rydberg atom chain.

\begin{figure}[htp!]
\centering  
\includegraphics[width=0.65\linewidth]{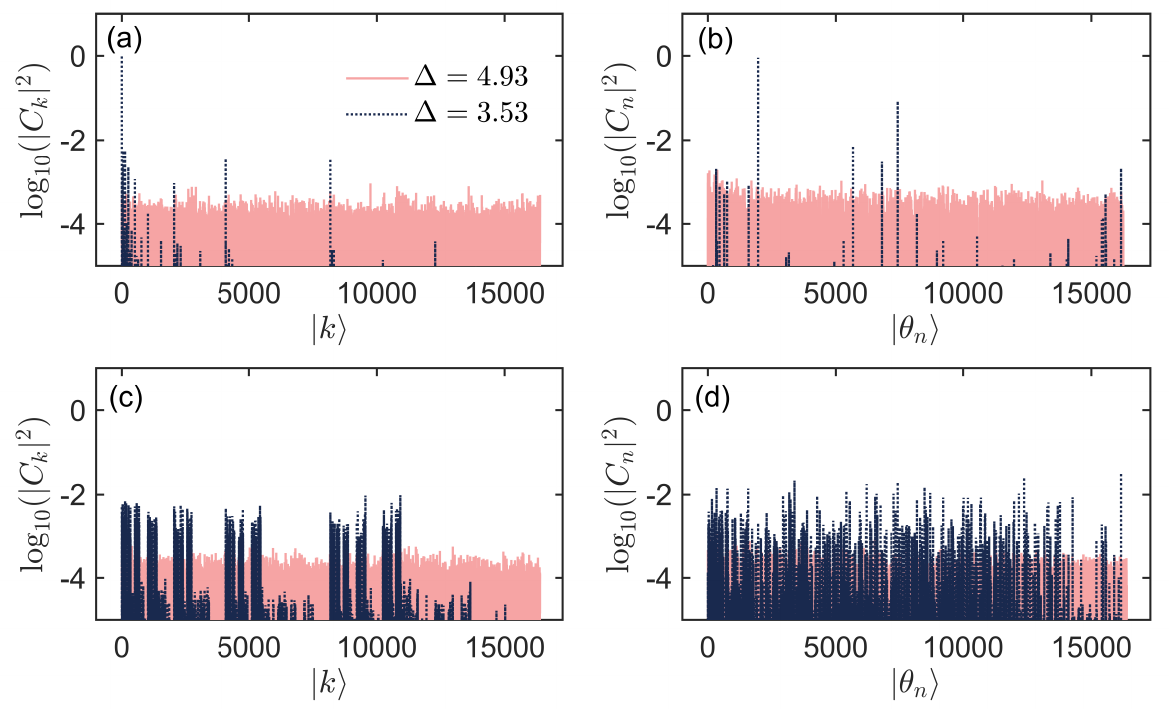}
\caption{Principal component analysis of long-time dynamics. The projections $|C_k|^2$ (left panels) and $|C_n|^2$ (right panels) for the final state $|\psi_t\rangle$ after $10^6$ driving periods, comparing the chaotic phase ($\Delta=4.93$, solid line) and the integrable phase ($\Delta=3.53$, dashed line). $C_k$ and $C_n$ are the projections onto the computational basis $|k\rangle$ and Floquet eigenstates $|\theta_n\rangle$, respectively. Panels (a,b) [(c,d)] present results obtained from the initial state $|\phi_0\rangle$ [$|\phi_1\rangle$]. The non-uniform, selective occupation in the integrable phase (dashed lines) confirms strong memory retention and GGE-like dynamics, contrasting sharply with the nearly uniform thermalized distribution in the chaotic phase (solid lines). Other parameters are $N=14$, $V_0=2$ and $\tau=\pi$. }
\label{N14principal_component}
\end{figure}

To demonstrate the distinct dynamical regimes at low driving frequencies (i.e., large $\tau$), we conduct a principal component analysis (PCA) of the final state $|\psi_t\rangle$~\cite{santos2012chaos}. We compute the projections of the final state onto both the computational basis states $|k\rangle$ ($C_k = \langle k | \psi_t \rangle$) and the Floquet eigenstates $|\theta_n\rangle$ ($C_n = \langle \theta_n | \psi_t \rangle$), after evolving the system for $10^6$ driving periods. The results are illustrated in Fig.~\ref{N14principal_component} showing a sharp distinction between the chaotic and integrable phases. 

The analysis of the computational basis projection, $|C_k|^2$, confirms the dynamics of local observables. For the initial state $|\phi_0\rangle$, the integrable phase exhibits strong localization, with the projection almost entirely concentrated on the ground state, whereas the chaotic phase distributes the projection among numerous states, signaling thermalization (Fig.~\ref{N14principal_component}(a)). Starting from the superposition state $|\phi_1\rangle$, the projection in the integrable phase remains non-uniform and localized within specific spectral bands, retaining the memory of the initial state, while the chaotic phase leads to a nearly uniform distribution across the entire basis (Fig.~\ref{N14principal_component}(c)). This non-uniformity in the integrable phase confirms the lack of thermalization of local observables. 

Crucially, the analysis in the Floquet eigenbasis, $|C_n|^2$, provides evidence for the nature of the athermal ensemble. In the chaotic phase, the projection onto the Floquet eigenstates is near-uniform across all eigenstates, consistent with the Eigenstate Thermalization Hypothesis (ETH) (Figs.~\ref{N14principal_component}(b,d)). However, for the integrable phase, the population is selectively confined to specific, non-uniformly occupied bands of eigenstates. This non-uniform, structured occupation strongly argues against the ETH and is the hallmark signature of a Generalized Gibbs Ensemble (GGE)-like ensemble or prethermalization in a driven system, where effective quasi-conserved quantities restrict the evolution to a low-dimensional subspace.

For real physical systems, one must also account for the finite atomic lifetime. As a concrete example, we consider the $|60S_{1/2}\rangle$ state of $^{87}\mathrm{Rb}$. Using the ARC package~\cite{robertson2021arc}, we find its lifetime to be approximately $230~\mu\mathrm{s}$ and obtain a vdW coefficient $C_6 \approx 139~\mathrm{GHz}\cdot(\mu\mathrm{m})^6$. Choosing a lattice spacing $a = 4.9~\mu\mathrm{m}$ yields an NN interaction strength $V_0 = {C_6}/{a^6} \approx 10~\mathrm{MHz}$. For a pulsed-laser Rabi frequency of $\Omega_0=5~\mathrm{MHz}$, all parameters may be expressed in units of $\Omega_0$, giving $\Omega_0 = 1, V_0/\Omega_0 = 2,\Omega_0\tau=\pi, \gamma/\Omega_0 \approx 8\times 10^{-4}$. The dynamics presented below are obtained by solving the master equation, $\dot{\rho}=-i[H,\rho]+\gamma \sum_j\left(\sigma_j^-\rho\sigma_j^+ -1/2\{\sigma_j^+\sigma_j^-,\rho\}\right)$, where $\sigma_j^-$ is the jump operator of the $j-$th atom. To obtain the continuous time evolution, we solve the master equation using the fourth-order Runge–Kutta algorithm. This is a heavy, time consuming calculation compared to the stroboscopic dynamics. Hence we have focused on relatively small systems ($N=10$ and $N=12$).

\begin{figure}[htp!]
\centering  
\includegraphics[width=0.7\linewidth]{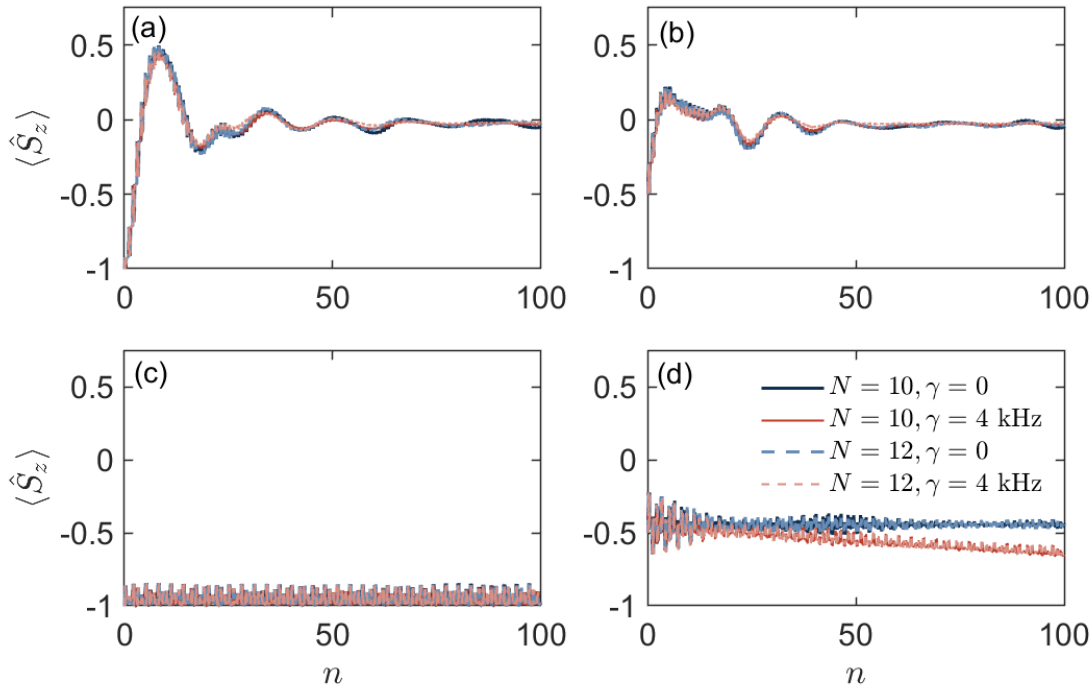}
\caption{Time evolution of the average population $\langle{\hat{S}_z\rangle}$ for systems without and with decay. (a,b) correspond to the chaotic phase ($\Delta/\Omega_0=4.93$) initialized in $|\phi_0\rangle$ (left) and $|\phi_1\rangle$ (right), respectively. (c,d) show the integrable phase ($\Delta/\Omega_0=3.53$) for the same initial states. The results are displayed for system sizes $N=10$ and $N=12$, with decay rates $\gamma=0$ and $\gamma=8\times 10^{-4}$ (realistic experimental value for $|60S_{1/2}\rangle$ state of $^{87}\mathrm{Rb}$ atom $\gamma=4~$kHz). Other parameters are $\Omega_0=5~$MHz, $V_0=10~$MHz and $\tau=0.2\pi~\mu$s.}
\label{decay case}
\end{figure}

Fig.~\ref{decay case} shows the time evolution of the average population $\langle \hat{S}_z \rangle$ in both the chaotic and integrable regimes, comparing cases with and without the atomic decay. In the chaotic phase ($\Delta = 4.93$), Figs.~\ref{decay case}(a,b) demonstrate that the dynamics are essentially insensitive to decay: for both initial states $|\phi_0\rangle$ and $|\phi_1\rangle$, the system exhibits damped oscillations and relaxes toward
$\langle \hat{S}_z \rangle \to 0$ within the simulated time window, consistent with the efficient thermalization in the continuous-time evolution. In contrast, the integrable phase ($\Delta = 3.53$) displays a markedly different response to decay. As shown in Fig.~\ref{decay case}(c), the initial state $|\phi_0\rangle$ remains close to the ground-state configuration throughout the evolution, and decay produces no qualitative change. However, for the initial state $|\phi_1\rangle$, Fig.~\ref{decay case}(d) illustrates that finite decay rate $\gamma$ induces a slow but noticeable drift away from the integrable trajectory. This drift results from the fact that the initial state is localized dynamically in the integrable phase. The Rydberg component of the state causes exponential decay. With this understanding in mind, we can still capture the localization feature in the dynamics.

\section{Effective Hamiltonian of the Floquet operator}

In one period $T=2\tau$, the many-body Hamiltonian can be written as,
 \begin{equation}\label{H}
 \left\{
 \begin{aligned}
 \hat{H}_1&=\sum_{j=1}^N\frac{\Omega_0}{2}\hat{\sigma}_j^x+\hat{H}_2,\\
 \hat{H}_2&=\sum_{j=1}^N \Delta \hat{n}_j+\sum_{j=1}^{N-1}\sum_{k>j}^{N}V_{jk}\hat{n}_j\hat{n}_{k},
 \end{aligned}
 \right.
 \end{equation}
where $\hat{H}_2$ is given by,
\begin{equation}\label{H_2_n}
\begin{aligned}
\hat{H}_2&=\frac{N\Delta}{2}+\frac{\Delta}{2}\sum_{j=1}^N\hat{\sigma}^z_j+\sum_{j=1}^{N-1}\sum_{k>j}^{N}\frac{V_{jk}}{4}\left(\hat{\mathbb{I}}+\hat{\sigma}^z_j+\hat{\sigma}^z_k+\hat{\sigma}^z_j\hat{\sigma}^z_k\right)\\
&=c-V_B(\hat{\sigma}^z_1+\hat{\sigma}^z_N)+\frac{1}{2}\sum_{j=1}^N \left(\Delta+V_0\left[1+\cdots+\frac{1}{(N-j)^6}\right]\right)\hat{\sigma}^z_j+\sum_{j=1}^{N-1}\sum_{k>j}^{N}\frac{V_{jk}}{4}\hat{\sigma}^z_j\hat{\sigma}^z_k,\\
\end{aligned}
\end{equation}
where the parameters $c$ and $V_B$ are defined as follows:
\begin{equation}
c=\frac{N\Delta}{2}+\frac{V_0}{4}\sum_j^{N-1}\frac{N-j}{j^6} = \frac{N\Delta}{2} + \frac{V_0}{4}[NL_6(N-1)-L_5(N-1)],
\end{equation}
and $V_B=\frac{V_0}{4}\sum_j^{N-2}\frac{1}{j^6} = \frac{V_0}{4}L_6(N-2)$. It is clear that $V_B$ is the strength of the detuning of the spin on the two edges of the chain. In the thermodynamic limit ($N\to\infty$), Hamiltonian $\hat{H}_2$ becomes,
\begin{equation}\label{H_2}
  \hat{H}_2=c-V_B(\hat{\sigma}^z_1+\hat{\sigma}^z_N)+\frac{1}{2} (\Delta+\frac{\pi^6}{945}V_0)\sum_{j=1}^N\hat{\sigma}^z_j+\sum_{j=1}^{N-1}\sum_{k>j}^{N}\frac{V_{jk}}{4}\hat{\sigma}^z_j\hat{\sigma}^z_k  .
\end{equation}

An effective Hamiltonian can be obtained by expanding the Floquet operator. The expansion normally converges in the integrable phase~\cite{kuwaharaFloquetMagnusTheory2016}. Thus, the effective Hamiltonian can be used to describe stroboscopic Floquet dynamics. To gain insight into the mechanism of thermalization control, we derive the effective Hamiltonian using the Baker–Campbell–Hausdorff (BCH) expansion~\cite{goldman2014periodically,kuwaharaFloquetMagnusTheory2016},
\begin{equation}\label{BCH}
\log(e^Xe^Y)=X+Y+\frac{1}{2}[X,Y]+\frac{1}{12}\left([X,[X,Y]]+[Y,[Y,X]]\right)+\cdots .
\end{equation}
This allows us to derive 
\begin{equation}\label{H_eff}
\begin{aligned}
\hat{H}_{F}&=\frac{i}{T}\log(e^{-i\hat{H}_2\tau}e^{-i\hat{H}_1\tau})\\
&\approx\underbrace{\frac{\hat{H}_2+\hat{H}_1}{2}}_{0-\text{order}}+\underbrace{\frac{-i\tau}{4}[\hat{H}_2,\hat{H}_1]}_{1-\text{order}}+\underbrace{\frac{-\tau^2}{24}\left([\hat{H}_2,[\hat{H}_2,\hat{H}_1]]+[\hat{H}_1,[\hat{H}_1,\hat{H}_2]]\right)}_{2-\text{order}} + \underbrace{\cdots}_{\text{Higher orders}}.\\
\end{aligned}
\end{equation}
In the above expansion, we have written the expansion up to the second-order explicitly.
The relevant commutation relations are given by,
\begin{equation}\label{1-order}
\begin{aligned}
&[\hat{H}_2,\hat{H}_1]=-i\Omega_0 V_B(\hat{\sigma}_1^y+\hat{\sigma}_N^y)+\frac{i\Omega_0}{2}(\Delta+\frac{\pi^6}{945}V_0)\sum_{j=1}^N\hat{\sigma}_j^y+\frac{i\Omega_0}{4}\sum_{j=1}^{N-1}\sum_{k>j}^{N}V_{jk}(\hat{\sigma}_j^y\hat{\sigma}_k^z+\hat{\sigma}_k^y\hat{\sigma}_j^z).\\ 
   & [\hat{H}_2,[\hat{H}_2,\hat{H}_1]]+[\hat{H}_1,[\hat{H}_1,\hat{H}_2]]=-\Omega_0^2V_{B}(\hat{\sigma}_1^z+\hat{\sigma}_N^z)+\frac{\Omega_0^2}{2}(\Delta+\frac{\pi^6}{945}V_0)\sum_{j=1}^N\hat{\sigma}_j^z+\frac{\Omega_0^2}{2}\sum_{j=1}^{N-1}\sum_{k>j}^{N}V_{jk}(\hat{\sigma}_j^z\hat{\sigma}_k^z-\hat{\sigma}_j^y\hat{\sigma}_k^y).
    \end{aligned}
\end{equation}
Substituting Eq.~(\ref{H_2}) and these commutators (\ref{1-order}) into Eq.~(\ref{H_eff}), we obtain the explicit form of the effective Hamiltonian,
\begin{equation}\label{H_eff_BCH}
\begin{aligned}	
\hat{H}_{F}&=c-\frac{\Omega_0\tau V_{B}}{4}(\hat{\sigma}_1^y+\hat{\sigma}_N^y)-\left(1-\frac{\Omega_0^2\tau^2}{24}\right)V_B(\hat{\sigma}_1^z+\hat{\sigma}_N^z)+\sum_{j=1}^N\left\{\frac{\Omega_0}{4}\hat{\sigma}_j^x+\frac{\Omega_0\tau}{8}(\Delta+\frac{\pi^6}{945}V_0)\hat{\sigma}_j^y\right.\\
&\left.+\left(\frac{1}{2}-\frac{\Omega_0^2\tau^2}{48}\right)(\Delta+\frac{\pi^6}{945}V_0)\hat{\sigma}_j^z\right\}+\sum_{j=1}^{N-1}\sum_{k>j}^{N}\frac{V_{jk}}{4}\left\{\left(1-\frac{\Omega_0^2\tau^2}{12}\right)\hat{\sigma}_j^z\hat{\sigma}_k^z+\frac{\Omega_0\tau}{4}(\hat{\sigma}_j^y\hat{\sigma}_k^z+\hat{\sigma}_k^y\hat{\sigma}_j^z)+\frac{\Omega_0^2\tau^2}{12}\hat{\sigma}_j^y\hat{\sigma}_k^y\right\}
\end{aligned}
\end{equation}
When $N\to\infty$, we may ignore the boundary terms and neglect the constant terms. Hamiltonian~(\ref{H_eff_BCH}) can be rewritten as follows,
\begin{equation}\label{Heff3}
	\begin{aligned}
		\hat{H}_{F}&=\sum_{j=1}^N\left\{ \frac{\Omega_0}{4}\hat{\sigma}_j^x+\frac{\Omega_0\tau}{8}(\Delta+\frac{\pi^6}{945}V_0)\hat{\sigma}_j^y+\left(\frac{1}{2}-\frac{\Omega_0^2\tau^2}{48}(\Delta+\frac{\pi^6}{945}V_0)\right)\hat{\sigma}_j^z\right\}\\
		&+\sum_{j=1}^{N-1}\sum_{k>j}^{N}\frac{V_{jk}}{4}\left\{\left(1-\frac{\Omega_0^2\tau^2}{12}\right)\hat{\sigma}_j^z\hat{\sigma}_{k}^z+\frac{\Omega_0\tau}{4}(\hat{\sigma}_j^y\hat{\sigma}_{k}^z+\hat{\sigma}_{k}^y\hat{\sigma}_j^z)+\frac{\Omega_0^2\tau^2}{12}\hat{\sigma}_j^y\hat{\sigma}_{k}^y\right\}\\
	\end{aligned}
\end{equation}
This expansion shows that the Rydberg atoms are coupled in three spin directions (i.e., the first line). The Rydberg atoms experience two-body vdW interactions along the $z$ and $y$ axes. The strength of the vdW interaction, on the other hand, is proportional to $\Omega_0^2\tau^2$. There is a long-range exchange interaction that is linearly proportional to $\Omega_0\tau$. In the fast-driven regime (i.e. $\tau\to 0$), the induced interaction may be neglected. The interaction parts mix spin couplings in different directions, which could lead to novel many-body physics.

\section{Free Fermions in the fast driven regime with $\Delta_0=0$.}
When $\Delta_{0}=0$, and parameter $\tau$ is small, the effective Hamiltonian can be further approximated to
\begin{equation}\label{Heff4}
\hat{H}_{F}\approx\sum_{j=1}^N \frac{\Omega_0}{4}\hat{\sigma}_j^x+\sum_{j=1}^{N-1}\frac{V_0}{4}\hat{\sigma}^z_j\hat{\sigma}^z_{j+1},
\end{equation}
where we have considered the NN interaction $V_0$ by letting $k=j+1$. This is a good approximation because the vdW interaction decays rapidly with increasing particle separation.

We use the Jordan–Wigner transformation to transform spin operators to fermion operators. First, we rotate our coordinate system around the y-axis mapping $\hat{\sigma}_x\rightarrow\hat{\sigma}_z$ and $\hat{\sigma}_z\rightarrow-\hat{\sigma}_x$.  Then the effective Hamiltonian becomes,
\begin{equation}\label{Heff_yaxis}		\hat{H}_{F}=\sum_j^N\frac{\Omega_0}{4}\hat{\sigma}_j^z+\sum_{j}^{N-1}\frac{V_0}{4}\hat{\sigma}_j^x\hat{\sigma}_{j+1}^x.\\
\end{equation}

The Jordan-Wigner transformation~\cite{jordan1928uber} is given by
\begin{equation}
	\left\{
	\begin{aligned}
		\hat{\sigma}_j^x&=\left\{\Pi_{m=1}^{j-1}(1-2c_m^\dagger c_m)\right\}(c_j^\dagger+ c_j),\\
		\hat{\sigma}_j^y&=-i\left\{\Pi_{m=1}^{j-1}(1-2c_m^\dagger c_m)\right\}(c_j^\dagger- c_j),\\
		\hat{\sigma}_j^z&=1-2c_j^\dagger c_j,\\
	\end{aligned}
	\right.
\end{equation}
where $c_j, c_j^\dagger$ are annihilation and creation operators for spinless fermions satisfying the anticommutation relations, i.e. $\{c_j, c_i^\dagger\}=\delta_{ji}, \{c_j, c_i\}=\{c_j^\dagger, c_i^\dagger\}=0$. Applying the Jordan–Wigner transformation, we can rewrite the Hamiltonian equation (\ref{Heff_yaxis}) with the fermionic operators,
\begin{equation}\label{Heff_JM}
	\begin{aligned}
		\hat{H}_{F}&\approx \sum_j^N\frac{\Omega_0}{4}(1-2c_j^\dagger c_j)+\sum_{j}^{N-1}\frac{V_0}{4}\left\{\Pi_{m=1}^{j-1}(1-2c_m^\dagger c_m)\right\}(c_j^\dagger+ c_j)\left\{\Pi_{n=1}^{j}(1-2c_n^\dagger c_n)\right\}(c_{j+1}^\dagger+ c_{j+1})\\
		&=\sum_j^N\frac{\Omega_0}{4}(1-2c_j^\dagger c_j)+\sum_{j}^{N-1}\frac{V_0}{4}(c_j^\dagger c_{j+1}^\dagger+c_{j+1}^{\dagger} c_j+c_j^\dagger c_{j+1}+ c_{j+1}c_j)\\
	\end{aligned}
\end{equation}

Next,  we introduce the collective (quasiparticle) operator $c_k, c_k^\dagger$ by performing Fourier transformation,
\begin{equation}\label{fourier}
\left\{
\begin{aligned}
c_j&=\frac{1}{\sqrt{N}}\sum_kc_ke^{-ikr_j},\\
c_j^\dagger&=\frac{1}{\sqrt{N}}\sum_kc_k^\dagger e^{ikr_j},\\
\end{aligned}
\right.
\end{equation}
The quasiparticle operator $c_k, c_k^\dagger$ are both satisfying the anticommutation relations $\{c_k, c_{k^\prime}^\dagger\}=\delta_{kk^\prime}$. Inserting Eq.~(\ref{fourier}) into Eq.~(\ref{Heff_JM}), this yields, 
\begin{equation}\label{Heff_fermion}
\begin{aligned}
    \hat{H}_{F}&\approx \frac{N\Omega_0}{4}+\sum_{k}\left\{\left(\frac{V_0(N-1)}{2N}\cos(ka)-\frac{\Omega_0}{2}\right)c_k^\dagger c_{k}+\frac{V_0(N-1)}{4N}\left(e^{-ika} c_k^{\dagger}c_{-k}^\dagger -e^{i{k}a}c_kc_{-k}\right)\right\}\\
    &=\sum_{k>0}\left\{\left(\frac{V_0(N-1)}{2N}\cos(ka)-\frac{\Omega_0}{2}\right)(c_k^\dagger c_{k}-c_{-k}c_{-k}^\dagger)-\frac{V_0(N-1)}{2N}i\sin(ka)\left (c_k^{\dagger}c_{-k}^\dagger-c_{-k}c_{k} \right)\right\}.\\
	\end{aligned}
\end{equation}

To diagonalize the Hamiltonian, we employ the Bogoliubov transformation, in which new fermion creation operators $\gamma_k$ are formed as a linear combination of $c_{-k}^\dagger$ and $c_{k}$ to remove terms in the Hamiltonian that do not conserve the particle number. These new operators are defined by a unitary transformation on the pair $\{c_k, c_{-k}^\dagger\}$,
\begin{equation}\label{Bogoliubov}
	\left(\begin{array}{c}c_k\\c_{-k}^{\dagger}\end{array}\right)
	=\left(\begin{array}{cc}u_k&iv_k\\iv_k&u_k\end{array} \right)	\left(\begin{array}{c}\gamma_k\\\gamma_{-k}^{\dagger}\end{array}\right)
\end{equation}
where $u_k, v_k$ are real numbers satisfying $u_k^2 + v_k^2 = 1, u_{-k} = u_k$, and $ v_{-k} =- v_k$. We can define $u_k = \cos(\theta_k), v_k= \sin(\theta_k)$. We rewrite the effective Hamiltonian in terms of the Bogoliubov fermions as follows
\begin{equation}\label{Heff_fermion_Bogo}
\begin{aligned}
   \hat{H}_{F} &\approx \sum_{k>0}\left\{\left[\frac{V_0(N-1)}{2N}\cos(2\theta_k+ka)-\frac{\Omega_0}{2}\cos(2\theta_k)\right]\gamma_{k}^{\dagger}\gamma_{k}\right.-\left[\frac{V_0(N-1)}{2N}\cos(2\theta_k+ka)-\frac{\Omega_0}{2}\cos(2\theta_k)\right]\gamma_{-k}\gamma_{-k}^{\dagger}\\
    &-i\left[\frac{V_0(N-1)}{2N}\sin(2\theta_k+ka)-\frac{\Omega_0}{2}\sin(2\theta_k)\right]\gamma_{k}^{\dagger}\gamma_{-k}^{\dagger}+\left.i\left[\frac{V_0(N-1)}{2N}\sin(2\theta_k+ka)-\frac{\Omega_0}{2}\sin(2\theta_k)\right]\gamma_{-k}\gamma_{k}\right\}\\
	\end{aligned}
\end{equation}

To diagonalize the Eq.~(\ref{Heff_fermion_Bogo}), we let the non-diagonal terms $\frac{V_0(N-1)}{2N}\sin(2\theta_k+ka)-\frac{\Omega_0}{2}\sin(2\theta_k)=0$, then $\tan(2\theta_k)=\frac{\sin(ka)}{h-\cos(ka)}$, where $h=\frac{\Omega_0 N}{V_0(N-1)}$. This leads to the energy of the Fermion,
\begin{equation}\label{Heff_eigen}
\epsilon_k=\frac{V_0(N-1)}{2N}\sqrt{1+h^2-2h\cos(ka)}.
\end{equation}
The final form of $\hat{H}_{F}$ becomes
\begin{equation}\label{H_eff_final}
\begin{aligned}
\hat{H}_{F}&\approx \sum_{k>0}\epsilon_k(\gamma_k^{\dagger}\gamma_k-\gamma_{-k}\gamma_{-k}^{\dagger}),\\
&=\sum_{k}\epsilon_k(\gamma_k^{\dagger}\gamma_k-\frac{1}{2}).\\
\end{aligned}
\end{equation}

\section{Bipartite entanglement entropy}

We calculate the half-chain entanglement entropy $\mathcal{S}_{N/2}/\langle{\mathcal{S}_{N/2}}\rangle$ as a function of the eigenphase $\theta_n$ for three different system sizes $N=10,12,$ and $14$. The eigenphases are unfolded within the interval $[0,2\pi)$ and are related to the quasienergies $\varepsilon_n$ through $\theta_n=\varepsilon_nT$,where $T$ is the driving period. Due to the discrete time-translation invariance of the Floquet dynamics, the spectrum is $2\pi$-periodic, such that $\theta_n' = \theta_n + 2\pi K$ (with $K$ an integer) corresponds to the same physical Floquet eigenstate. This eigenphase periodicity is a key ingredient for understanding Floquet thermalization and the structure of Floquet eigenstates discussed below.

\begin{figure}[htp!]
\centering  
\includegraphics[width=0.98\linewidth]{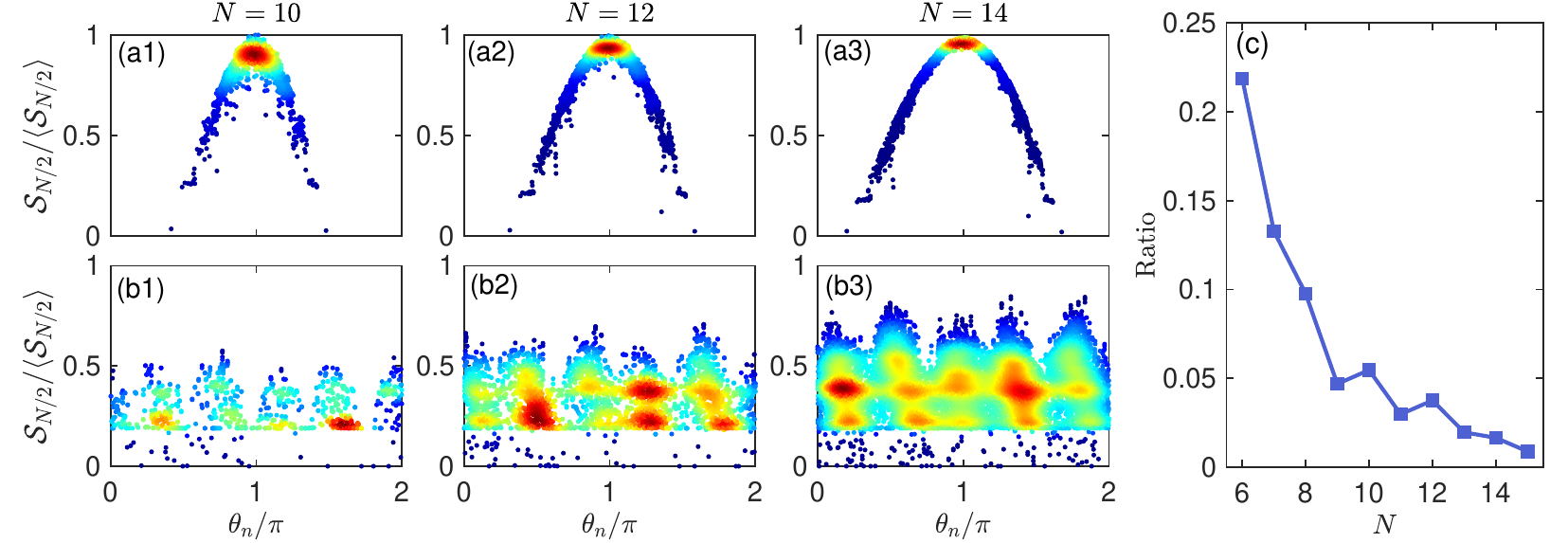}
\caption{The entangle entropy of half chain $\mathcal{S}_{N/2}/\langle\mathcal{S}_{N/2}\rangle$ as a function of Floquet quasienergy phase $\theta_n$ for (a1-a3) the chaotic phase $\Delta=4.93$ and (b1-b3) the integrable phase $\Delta=3.53$ with different atom number $N=10,12,$ and $14$. The color scheme indicates a higher density of states for warmer colors. (c) The ratio of eigenstates whose entanglement entropy is below half of the maximum value ($\mathcal{S}_{N/2}/\langle\mathcal{S}_{N/2}\rangle<0.5$) as a function of system size $N$. Other parameter are $V_0=2$ and $\tau=\pi$.}
\label{theta_entropy_N10_12_14}
\end{figure}

The color scheme reflects the normalized local density of states, with warmer colors corresponding to higher local densities. In the chaotic phase, the half-chain entanglement entropy exhibits a continuous, unimodal distribution peaking near $\theta_n \approx \pi$, where the quasienergy density is also concentrated. This indicates that the eigenstates explore the full Hilbert space, which is consistent with the quantum thermalization hypothesis. Importantly, it can be seen that many eigenstates have high entropy (the red area). As the system size $N$ increases, the distribution converges toward the Page value, reflecting the thermal behavior in the thermodynamic limit [see Figs.~\ref{theta_entropy_N10_12_14}(a1–a3)].

In contrast, the integrable phase exhibits a discrete entropy distribution with clustered quasienergies [see Figs.~\ref{theta_entropy_N10_12_14}(b1–b3)]. The entanglement of the eigenstate is confined to the low entry region ($0\sim 0.5\langle \mathcal{S}_{N/2} \rangle$), reflecting a nonthermal behavior constrained by conserved quantities. Although the distribution becomes denser with a larger $N$, it retains a clear discreteness, which is a characteristic of integrable systems. These results confirm the chaotic behavior at $\Delta = 4.93$ and the integrable phase at $\Delta = 3.53$, consistent with the analysis of the level spacing statistics, inverse participation ratio, and stroboscopic dynamics.

We find that there exist low-entanglement states at the edges of the spectrum in the chaotic phase. To further clarify this, we calculate the fraction of Floquet eigenstates whose entanglement entropy is below half of the maximum possible value ($\mathcal{S}_{N/2}/\langle\mathcal{S}_{N/2}\rangle<0.5$) as a function of system size $N$. As shown in the Fig.~\ref{theta_entropy_N10_12_14}(c), this ratio decreases rapidly with increasing $N$, indicating that the low-entanglement states observed in small systems are finite-size effects. For $N=15$, the ratio is already less than $1\%$, suggesting that these remaining low-entanglement states may correspond to quantum many-body scar states, which are of intrinsic interest and warrant further investigation.

\section{Parity and degenerate eigenstates}\label{degeneracy discussion}
Hamiltonian Eq.~(1) in the main text is invariant under reflection. In other words, $\hat{H}$ commutes with the parity operator~\cite{joel2013introduction}
\begin{equation}\label{parity}
\hat{\Pi}=\left\{
\begin{aligned}
\hat{\mathcal{P}}_{1,L}\hat{\mathcal{P}}_{2,L-1}...\hat{\mathcal{P}}_{\frac{L}{2},\frac{L+2}{2}},\quad&\text{for}\quad L=\text{even}\\
\hat{\mathcal{P}}_{1,L}\hat{\mathcal{P}}_{2,L-1}...\hat{\mathcal{P}}_{\frac{L-1}{2},\frac{L+3}{2}},\quad&\text{for}\quad L=\text{odd}
\end{aligned}
\right.
\end{equation}
where $\hat{\mathcal{P}}_{i,j}=(\hat{\sigma}_i^x\hat{\sigma}_j^x+\hat{\sigma}_i^y\hat{\sigma}_j^y+\hat{\sigma}_i^z\hat{\sigma}_j^z+\hat{\mathbb{I}})/2$ is the permutation operator and $\hat{\mathbb{I}}$ is the identity operator.

If $[\hat{H}, \hat{\Pi}] = 0$, then a non-degenerate eigenstate $\ket{\psi_j}$ of $\hat{H}$ is also an eigenstate of the parity operator $\hat{\Pi}$, i.e., $\hat{H}\ket{\psi_j}=E_j\ket{\psi_j}$ and $\hat{\Pi}\ket{\psi_j}=\Pi\ket{\psi_j}$, where $\Pi=\pm1$. In our calculations, we encounter degenerate eigenstates $\ket{\phi_m}$ that do not have a well-defined parity in general, $\hat{\Pi}\ket{\phi_m}\not=\pm\ket{\phi_m}$. In the degenerate subspace, we can define $\ket{\phi}=\sum_{sub}c_m\ket{\phi_m}$, where $\braket{\phi}{\phi}=1$, and $c_m$ are coefficients. Then we have $\hat{H}\ket{\phi}=E_\lambda\ket{\phi}$, where $E_\lambda$ is the degenerate eigenvalue. But due to the existence of coefficients $c_m$, $\ket{\phi}$ cannot have a well-defined parity even when $\ket{\phi_m}$ has. It is possible to obtain a well-defined parity in the degenerate subspace. Considering the double degeneracy case, $\ket{\phi_1}$ and $\ket{\phi_2}$, the parity operator $\hat{\Pi}$ is constructed as,
\begin{equation}
\hat{\Pi}=
\left(
\begin{array}{cc}
\bra{\phi_1}\hat{\Pi}\ket{\phi_1}  &\bra{\phi_1}\hat{\Pi}\ket{\phi_2} \\\bra{\phi_2}\hat{\Pi}\ket{\phi_1}
&\bra{\phi_2}\hat{\Pi}\ket{\phi_2} 
\end{array}
\right).
\end{equation}
This leads to eigenstate $\ket{+}=b_1\ket{\phi_1}+b_2\ket{\phi_2}$ and $\ket{-}=c_1\ket{\phi_1}+c_2\ket{\phi_2}$, where $\ket{\pm}$ have even and odd parity with eigenenergy $\hat{H}\ket{\pm}=E_\lambda\ket{\pm}$. $b_j$ and $c_j$ are normalization coefficients. For multiple degeneracies, one can extend to a higher dimension and construct eigenstates that preserve the parity. In our calculations, we have found several groups of double degenerate eigenstates. We use the above method to treat each group individually.

\section{Classical spins dynamics}
The Rydberg atom chain with open boundary conditions and NN interaction is described by the Hamiltonian
\begin{equation}
\hat{H}(t)=\left\{
\begin{aligned}
\hat{H}_1&=\sum_j^N\frac{\Omega_0}{2}\hat{\sigma}_j^x+\hat{H}_2,\quad t\in[0,\tau]\\
\hat{H}_2&=\sum_j^N\frac{\Delta+V_0}{2} \hat{\sigma}_j^z+\sum_{j=1}^{N-1}\frac{V_{0}}{4}\hat{\sigma}_j^z\hat{\sigma}_{j+1}^z-\frac{V_{0}}{4}(\hat{\sigma}_1^z+\hat{\sigma}_N^z),\quad t\in[\tau,T]
\end{aligned}
\right.
\end{equation}

In the classical limit, the time evolution of the system is governed by Hamilton’s equation of motion (EOM) $\dot{\hat{\sigma}}_j^\mu(t)=\{\hat{\sigma}_j^\mu, \hat{H}(t)\}$. The stroboscopic dynamics can be obtained by a discrete map $\vec{\sigma}_j(nT)=[\tau_2\circ\tau_1]^n[\vec{\sigma}_j(0)]$, with $n\in \mathcal{N}$ counting the driving cycles~\cite{howell2019asymptotic}. The map $\tau_2\circ\tau_1$ is a classical analog of the quantum Floquet unitary. For the first half-period, the system follows the nonlinear rotation matrix $\tau_1$, which is given by
\begin{equation}\label{tau1}
\tau_1(\vec{\sigma}_j) = 
\begin{bmatrix}
\hat{\sigma}_j^x \frac{\Omega_0^2 + \alpha_j^2 \cos(\omega_0 t)}{D}
+ \hat{\sigma}_j^y \left( -\frac{\alpha_j}{\sqrt{D}} \sin(\omega_0 t) \right)
+ \hat{\sigma}_j^z \frac{\Omega_0 \alpha_j (1 - \cos(\omega_0 t))}{D} \\[6pt]
\hat{\sigma}_j^x \frac{\alpha_j}{\sqrt{D}} \sin(\omega_0 t)
+ \hat{\sigma}_j^y \cos(\omega_0 t)
+ \hat{\sigma}_j^z \left( -\frac{\Omega_0}{\sqrt{D}} \sin(\omega_0 t) \right) \\[6pt]
\hat{\sigma}_j^x \frac{\Omega_0 \alpha_j (1 - \cos(\omega_0 t))}{D}
+ \hat{\sigma}_j^y \frac{\Omega_0}{\sqrt{D}} \sin(\omega_0 t)
+ \hat{\sigma}_j^z \frac{\alpha_j^2 + \Omega_0^2 \cos(\omega_0 t)}{D}
\end{bmatrix},
\end{equation}
with spin-dependent frequency of rotation $\alpha_j=\Delta+V_0+V_0/2(\hat{\sigma}_{j-1}^z+\hat{\sigma}_{j+1}^z)$ for $j\in[2,N-1]$, and $\alpha_{1(N)}=\Delta+V_0/2+V_0/2\hat{\sigma}_{2(N-1)}^z$ for boundary term. The parameters $D=\alpha_j^2+\Omega_0^2,\omega_0=\sqrt{D}/2$. Similarly, in the second half-period the rotation matrix $\tau_2$ is given by
\begin{equation}\label{tau2}
	\tau_2(\vec{\sigma}_j) = 
	\begin{bmatrix}
		\hat{\sigma}_j^x \cos(\alpha_j t/2)  - \hat{\sigma}_j^y \sin(\alpha_j t/2) \\[6pt]
		\hat{\sigma}_j^x \sin(\alpha_j t/2)  + \hat{\sigma}_j^y \cos(\alpha_j t/2) \\[6pt]
		\hat{\sigma}_j^z
	\end{bmatrix}.
\end{equation}

\begin{figure}[htp!]
	\centering  
	\includegraphics[width=0.5\linewidth]{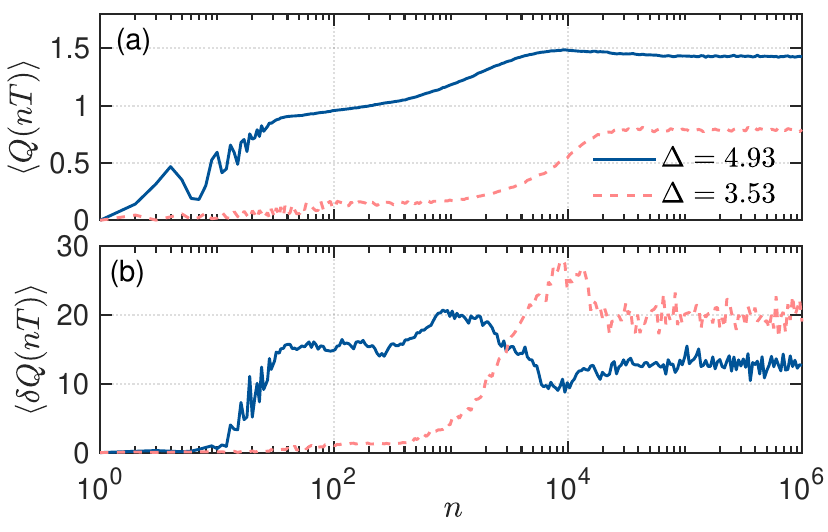}
	\caption{Noise-averaged energy (a) and energy variance (b) over 100 noisy initial-state realizations as a function of the number of driving cycles $n$ for the chaotic phase $\Delta=4.93$ and the integrable phase $\Delta=3.53$. The chaotic regime heats rapidly and saturates, while the integrable regime shows weaker energy growth but larger long-time fluctuations. Other parameters are $N=100$, $V_0=2$ and $\tau=\pi$. }
	\label{Noise-averaged energy}
\end{figure}

We prepare the initial state of the system in the ground state (GS) of the time-averaged Hamiltonian, $\hat{H}_{a}=(\hat{H}_{1}+\hat{H}_{2})/2=\sum_j^N\frac{\Omega_0}{4}\hat{\sigma}_j^x+\sum_j^N\frac{\Delta+V_0}{2} \hat{\sigma}_j^z+\sum_{j=1}^{N-1}\frac{V_{0}}{4}\hat{\sigma}_j^z\hat{\sigma}_{j+1}^z-\frac{V_{0}}{4}(\hat{\sigma}_1^z+\hat{\sigma}_N^z)$, which is confined to the $xz$-plane. The optimal azimuthal angle for each spin is determined by minimizing the system energy. To study the thermalization behavior from a GS configuration, we introduce weak random perturbations to the azimuthal angle of each spin, sampled from a uniform distribution over the interval $[-100/\pi,100/\pi]$. The perturbation amplitude is chosen to be sufficiently small such that it does not alter the underlying dynamical behavior. The following quantities—the normalized energy and its variance—are evaluated as averages over $100$ independent realizations of this initial noise:
\begin{align}
\langle Q(nT) \rangle &= \frac{\langle \hat{H}_{a}[\{\vec{\sigma}_j(nT)\}] \rangle - E_{\mathrm{GS}}}
{- E_{\mathrm{GS}}}, \\[6pt]
\langle \delta Q(nT) \rangle &= 
\sqrt{\langle \hat{H}_{a}^{2}[\{\vec{\sigma}_j(nT)\}] \rangle - \langle \hat{H}_{a}[\{\vec{\sigma}_j(nT)\}] \rangle^{2}}.
\end{align}

The dynamical evolution obtained from the classical model is shown in Fig.~\ref{Noise-averaged energy}. For the chaotic phase ($\Delta=4.93$), the energy increases rapidly during the early driving cycles and then gradually approaches a saturation plateau, which is consistent with classical heating toward a steady state. The corresponding energy variance remains relatively small and does not exhibit strong late-time growth, indicating that classical trajectories are only weakly sensitive to noise. In contrast, the integrable phase ($\Delta=3.53$) displays a much slower but still noticeable increase in energy. However, its variance grows substantially and remains large over a broad time window, reflecting a strong sensitivity to initial-state noise and the presence of long-lived correlations typical of near-integrable dynamics.
